\documentclass{jpsj2}

\newcommand{\rev}[1]{#1}

\newcommand{\Prob}{\mbox{Prob}}
\newcommand{\sgn}{\mbox{sgn}}
\newcommand{\erf}{\mbox{erf}}
\renewcommand{\vec}[1]{\mbox{\boldmath $#1$}}

\title{Stochastic transitions of attractors in associative memory
models with correlated noise}

\author{Masaki \textsc{Kawamura}$^{1}$
\thanks{E-mail address: kawamura@sci.yamaguchi-u.ac.jp}
and 
Masato \textsc{Okada}$^{2}$$^{3}$ 
}

\inst{$^{1}$Graduate School of Science and Engineering, 
 Yamaguchi University \\
 Yoshida 1677-1, Yamaguchi-shi, Yamaguchi, 753-8512 Japan \\
$^{2}$Department of Complexity Science and Engineering,
 Graduate School of Frontier Sciences, The University of Tokyo, 
 5-1-5 Kashiwanoha, Kashiwa-shi, Chiba, 277-8562 Japan, \\
$^{3}$Computational Neuroscience Research Group, RIKEN Brain Science
Institute, 2-1 Hirosawa, Wako, Saitama, 351-0198 Japan.
}

\abst{We \rev{investigate} dynamics of recurrent neural networks with
correlated noise to analyze the noise's effect.  The mechanism of
correlated firing has been analyzed in various models, but its
functional roles have not been discussed in sufficient detail. Aoyagi
and Aoki have shown that the state transition of a network is invoked by
synchronous spikes.
 We introduce two types of noise to each neuron:
 thermal independent noise and correlated noise.  Due to the effects
 of correlated noise, the correlation between neural inputs cannot
 be ignored, so the behavior of the network has sample dependence.
 We discuss two types of associative memory models: one with auto- and
 weak cross-correlation connections and one with hierarchically
 correlated patterns.  The former is similar in structure to Aoyagi and
 Aoki's model. We show that stochastic transition can be presented by
 correlated rather than thermal noise. In the latter, we show stochastic
 transition from a memory state to a mixture state using correlated
 noise.  To analyze the stochastic transitions, we derive a macroscopic
 dynamic description as a recurrence relation form of a probability
 density function when the correlated noise exists. Computer simulations
 agree with theoretical results.
}

\kword{common external input, correlated noise, associative memory,
stochastic transition, macroscopic dynamic description}

\begin{document}
\maketitle

\section{Introduction} 

In the activities of nerve cells, synfire chains, i.e., synchronous
firings of neurons, can often be observed \cite{Abeles1991}. To analyze
the mechanism of synchronous firings, conditions for propagating them
between layers have been investigated in layered neural networks
\cite{Diesmann_etal1999,CateauFukai2001}. Common synaptic inputs to
neurons have been introduced in layered neural networks, and correlated
firings of neurons have been investigated on a theoretical level
\cite{Amari_etal2003}. We have also analyzed common synaptic inputs
in recurrent neural networks, i.e., a sequential associative memory model
\cite{Kawamura2005}, and succeeded in deriving a macroscopic dynamic
description as a recurrence relation form of a probability density
function.
On another front, investigating the functional roles of synchronous
firings has become important. Associative memory models, which store
patterns as attractors, are used to explain these roles. Aoyagi and Aoki
\cite{AoyagiAoki2004,AokiAoyagi2006} have investigated a model with
auto- and weak cross-correlation connections and showed that a
transition between attractors cannot be invoked by thermal independent
noise but can be by synchronous spikes.  In general, when using thermal
independent noise, the state leaves an attractor for another
one. However, Aoyagi and Aoki's model can make the state transit to the
next attractor by synchronous spikes, i.e., external inputs.

In this paper, we consider two types of associative memory models with
{\itshape a common external input}.  The common external input affects
all neurons equally and therefore invokes correlated firings of
neurons. To analyze dynamic behavior theoretically, we assume that
common external input obeys Gaussian distribution and acts
like correlated, not independent, noises.

\begin{figure}[tb]
 \hfill (a)~\includegraphics[width=65mm]{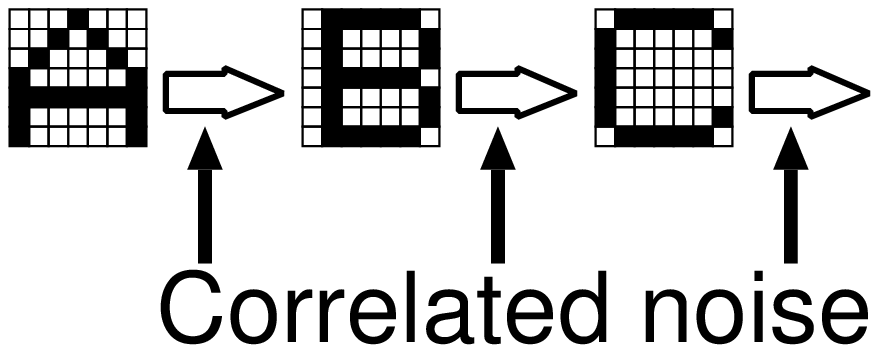}
 \hfill (b)~\includegraphics[width=45mm]{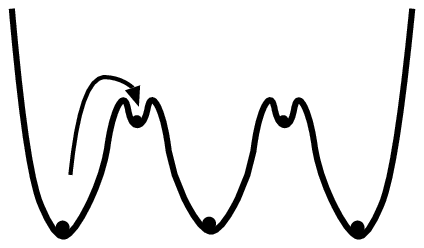}
 \hfill \mbox{}

 \caption{\rev{Schematic illustrations of }
 (a) stochastic transition by correlated noise \rev{in an associative
 memory model with auto- and weak cross-correlation connections, and }
 (b) transition from memory to mixture state
 \rev{in one with hierarchically correlated patterns. } 
 \label{fig:concept}}
\end{figure}

The first model consists of an autoassociative memory model
\cite{Hopfield1982,Okada1995} and a weakly connected sequential
associative memory model
\cite{Amari1988b,KitanoAoyagi1998,Katayama2001,Kawamura2002}, as well as
Aoyagi and Aoki's model \cite{AoyagiAoki2004,AokiAoyagi2006}. Associative
processing requires functions of both autoassociative memory, which
retrieves a memory pattern from an ambiguous initial state, and
sequential associative memory, which retrieves memory patterns
episodically \cite{Kawamura1999}. This model contains these
functions. To switch from autoassociative to sequential associative, we
introduce external inputs.  One input is an independent (thermal) noise
and the other is a correlated noise (a common external input).  The
synchronous spikes in Aoyagi and Aoki's model can be considered
correlated noises in our model since they invoke correlated firings. We
examine the effectiveness of the independent and correlated noises, and
demonstrate that stochastic transition can be presented by correlated,
but not independent, noise, as shown in Figure~\ref{fig:concept}(a).

The second model we consider is an associative memory model that stores 
hierarchically correlated patterns
\cite{Amari1977,PargaVirasoro1986,Gutfreund1988,KakeyaKindo1996,Toya2000}.
In this case, both the memory states and their mixture states are
attractors \cite{AmitGutfreund1985}. We demonstrate that a stochastic
transition from a memory to a mixture state
can be invoked by the correlated rather than the independent noise,
as shown in Figure~\ref{fig:concept}(b). 

Almost all existing models have been analyzed by applying independent
units or neurons at the thermodynamic limit. Since we introduce
correlated noises in these models, the sum of the inputs to the neurons
are correlated. Therefore, the firings of the neurons are also
correlated. With an infinite number of neurons, there is no
thermodynamic limit and sample dependence appears
\cite{Amari_etal2003,Sakai2002}.  Theoretical approaches that address
sample dependence are described by probability density functions (PDF)
for macroscopic states or order parameters
\cite{YamanaOkada2005,Kawamura2005}.  Yamana and Okada
\cite{YamanaOkada2005} have introduced uniform common synaptic inputs
that depend on preneurons to a layered associative memory model and
derived the PDF. Kawamura {\itshape et al.}  \cite{Kawamura2005} have
introduced common synaptic inputs to recurrent neural networks and
derived a macroscopic dynamic description as a recurrence relation form
of the PDF. The common synaptic inputs that depend on preneurons were
analyzed in these previous papers
\cite{Peretto1988,YoshiokaShiino1997,Kappen2005}. In this paper, we
analyze external rather than synaptic inputs and derive a macroscopic
dynamic description as a recurrence relation form of the PDF.

\rev{This paper is organized as follows. In the next section, we examine
the first model, i.e., the model with auto- and weak cross-correlation
connections.  Theoretical results and those obtained using computer
simulations are given. In \S\ref{sec:hierar}, we examine the second
model, i.e., the model that stores hierarchically correlated patterns.
The final section is devoted to conclusion.
}

\section{Autoassociative and sequential associative memory model}
\label{sec:auto-seq}

\subsection{Model}

We consider the associative memory model consisting of $N$ units or
neurons and introduce two types of external input: independent
external input $\zeta_i^t$, which affects each neuron independently, and
common external input $\eta^t$, which affects all neurons mutually. The
state of the neurons takes $x^t_i=\pm1$ and is updated synchronously by
\begin{equation}
 x_i^{t+1} = F\left(\sum_{j\neq i}^NJ_{ij}x_j^t+\zeta_i^t+\eta^t\right),
  \label{eqn:dynamics}
\end{equation}
where the output function is $F(h)=\sgn(h)$, and $J_{ij}$ is a synaptic
connection from the $j$th to the $i$th neuron.  Here, the synapse has
both auto- and weak cross-correlation connections and stores $p$ random
patterns $\vec{\xi}^{\mu}=(\xi^{\mu}_1,\cdots,\xi^{\mu}_N)^T$,
$\mu=1,2,\cdots,p$. The element of the patterns takes $\pm1$ with
\begin{equation}
 \Prob\left[\xi^{\mu}_i=\pm1\right]=\frac{1}{2}.
\end{equation}
The synaptic connection $J_{ij}$ is given by 
\begin{equation}
 J_{ij}=\frac1N\sum_{\mu=1}^{p}\xi^{\mu}_i\xi^{\mu}_j
  +\frac{\varepsilon}{N}\sum_{\mu=1}^{p}\xi^{\mu+1}_i\xi^{\mu}_j,
  \label{eqn:Jij}
\end{equation}
where $\xi^{p+1}_i=\xi^1_i$ and $\varepsilon$ denotes a connecting
parameter. Under normal conditions, that is, without noise, the model
behaves as an autoassociative memory model, since the connecting
parameter is very weak $\varepsilon\ll1$.

We define the overlap by the direction cosine between the state of
neurons, $\vec{x}^t$, at time $t$ and the memory pattern
$\vec{\xi}^{\mu}$,
\begin{equation}
 m^{\mu}_t = \frac{1}{N}\sum_{i=1}^N \xi_i^{\mu} x_i^t .
  \label{eqn:mt}
\end{equation}
From (\ref{eqn:Jij}) and (\ref{eqn:mt}), the state of the neurons,
$x_i^{t+1}$, becomes
\begin{eqnarray}
 x_i^{t+1} &=&
  F \left(\sum_{\mu=1}^{p}\xi^{\mu}_im^{\mu}_t
     +\varepsilon\sum_{\mu=1}^{p}\xi^{\mu+1}_im^{\mu}_t
     +\zeta_i^t+\eta^t \right)  \\
 &=& F \left(\sum_{\mu=1}^{p}
        \left(\xi^{\mu}_i+\varepsilon\xi^{\mu+1}_i\right)m^{\mu}_t
	+\zeta_i^t+\eta^t \right) .
 \label{eqn:x_xi}
\end{eqnarray}
The initial state $\vec{x}^0$ is determined according to the probability
distribution
\begin{equation}
 \Prob[x_i^0=\pm1] = \frac{1\pm m_0 \xi^{1}_i}{2}.
\end{equation}
Therefore, the overlap between the pattern $\vec{\xi}^1$ and the initial
state $\vec{x}^0$ is $m_0$.  

\subsection{Theory}

Let us derive the macroscopic description of the model with the external
inputs. In order to simplify the argument, we assume that the number of
memory patterns, $p$, is finite, especially when $p=3$. The independent
external input $\zeta_i^t$ is time independent and independent of each
neuron $x^t_i$. In addition, we assume $\zeta_i^t$ obeys the Gaussian
distribution with ${\cal N}\left(0, \Delta^2\right)$. The correlated
external input $\eta^t$ is time independent, and is assumed to obey the
Gaussian distribution with ${\cal N}\left(0, \delta^2\right)$.  These
external inputs are called independent noise and correlated noise.

\subsubsection{Without correlated noise ($\eta^t=0$)}

First, we discuss the case without correlated noise, $\eta^t=0$.
Since there is only independent noise, when $N\to\infty$, the overlap
$m^{\mu}_{t+1}, \mu=1,2,3$ becomes
\begin{eqnarray}
 m^{\mu}_{t+1} &=& \frac{1}{N}\sum_{i=1}^N \xi^{\mu}_i
  F\left(\sum_{\nu=1}^{3}
    \left(\xi^{\nu}_i+\varepsilon\xi^{\nu+1}_i\right)m^{\nu}_t
    +\zeta^t_i\right), \\
 &=& \! \left< \! \int \!\! D_z \xi^{\mu}
      F\left(\sum_{\nu=1}^{3}
        \left(\xi^{\nu}+\varepsilon\xi^{\nu+1}\right)m^{\nu}_t
        +\Delta z\right)\!\! \right>_{\xi} \!\! , \\
 &=& \left< \xi^{\mu}
      \erf\left(\frac{\sum_{\nu=1}^{3}
        \left(\xi^{\nu}+\varepsilon\xi^{\nu+1}\right)m^{\nu}_t }
           {\sqrt{2}\Delta}\right)\right>_{\xi} ,
      \label{eqn:merf}
\end{eqnarray}
where $D_{z}=\frac{dz}{\sqrt{2\pi}} \exp\left(-\frac{z^2}{2}\right) $ and
$\left<\cdot\right>_{\xi}$ denotes the average over the memory pattern
$\vec{\xi}^{\mu}$. The function $\erf(u)$ is defined as
\begin{equation}
 \erf\left(u\right)=\frac{2}{\sqrt{\pi}}\int_0^xdt \exp\left(-t^2\right).
\end{equation}
In this case, we find that the macroscopic parameter $m^{\mu}_t$ is
deterministically given.

\subsubsection{With correlated noise ($\eta^t\neq0$)}

In the case with correlated noise, the states of the neurons are
correlated, because the common external input is introduced into all
neurons mutually. Therefore, sample dependence must be taken into
account.  We derive a macroscopic dynamic description as a recurrence
relation form of a probability density function (PDF).

The correlated noise distributes $\eta^t\sim {\cal
N}\left(0,\delta^2\right)$. When $\eta^t$ is known at
given time $t$ and $N\to\infty$, then $m^{\mu}_{t+1}, \mu=1,2,3$ can
be given as a function of $m^{1}_t$, $m^{2}_t$, $m^{3}_t$, and $\eta^t$,
\begin{eqnarray}
 m^{\mu}_{t+1}(m^{1}_t,m^{2}_t,m^{3}_t,\eta^t)
 &=& \left<\int D_{z}
      \xi^{\mu} F\left(\sum_{\nu=1}^{3}
                  \left(\xi^{\nu}+\varepsilon\xi^{\nu+1}\right)m^{\nu}_t
                  +\Delta z+\eta^t\right) \right>_{\xi},  \\
 &=& \left< \xi^{\mu}
      \erf\left(\frac{\sum_{\nu=1}^{3}
        \left(\xi^{\nu}+\varepsilon\xi^{\nu+1}\right)m^{\nu}_t+\eta^t}
           {\sqrt{2}\Delta}\right)\right>_{\xi} .
\end{eqnarray}
Using this equation, the dynamic behavior of the model for various
$\delta$ can be analyzed. When $\delta=0$, the model corresponds to the
existing associative memory models and behaves deterministically.
However, when $\delta>0$, the values of $m^{1}_t$, $m^{2}_t$, and
$m^{3}_t$ are distributed. This distribution can be described as the
PDF, $p\left(m^{1}_t,m^{2}_t,m^{3}_t,\eta^t\right)$. Since $\eta^t$ is
independent of $m^{1}_t$, $m^{2}_t$, and $m^{3}_t$,
$p\left(m^{1}_t,m^{2}_t,m^{3}_t,\eta^t\right)$ can be divided into two
PDFs:
\begin{equation}
 p\left(m^{1}_t,m^{2}_t,m^{3}_t,\eta^t\right) =
  p\left(m^{1}_t,m^{2}_t,m^{3}_t\right)p\left(\eta^t\right) .
  \label{eqn:independ}
\end{equation}
Therefore, the PDF can be given by
\begin{eqnarray}
 p\left(m^{1}_{t+1},m^{2}_{t+1},m^{3}_{t+1}\right)
  &=& \int \prod_{\nu=1}^3 dm^{\nu}_t d\eta^t p\left(m^{1}_t,m^{2}_t,m^{3}_t\right)
  p\left(\eta^t\right) \nonumber \\
 &\times& \prod_{\nu=1}^3
  \delta\left(m^{\nu}_{t+1}-m^{\nu}_{t+1}(m^{1}_t,m^{2}_t,m^{3}_t,\eta^t)\right),
  \label{eqn:Pnext}
\end{eqnarray}
where $\delta(\cdot)$ denotes the Dirac delta function defined as
\begin{equation}
 \delta\left(u\right)=
  \left\{\begin{array}{ll}
   \rev{\infty}, & u=0 \\
   0, & u\neq0 \\
	 \end{array}\right. .
\end{equation}
The PDF $p\left(\eta^t\right)$ is given by
\begin{equation}
 p\left(\eta^t\right) = \frac{1}{\sqrt{2\pi}\delta}
  \exp\left(-\frac{\left(\eta^t\right)^2}{2\delta^2}\right).
\end{equation}
Then, we combine this with the integral of $\eta^t$ and obtain equations
\begin{eqnarray}
 p\left(m^{1}_{t+1},m^{2}_{t+1},m^{3}_{t+1}\right)
 &=&  \int \! \prod_{\nu=1}^3dm^{\nu}_t
  p\left(m^{1}_t,m^{2}_t,m^{3}_t\right)
  K\left(m^{\nu}_{t+1};m^{\nu}_t\right) ,
 \label{eqn:PnextK} \\
 K\left(m^{\nu}_{t+1};m^{\nu}_t\right)
  &=&\int \! d\eta^tp\left(\eta^t\right)
  \prod_{\nu=1}^3
  \delta\left(m^{\nu}_{t+1}-m^{\nu}_{t+1}(m^{1}_t,m^{2}_t,m^{3}_t,\eta^t)\right).
  \label{eqn:Kernel}
\end{eqnarray}
We can also calculate (\ref{eqn:Kernel}) directly, but we evaluate it
using the Monte Carlo method in the following section.

\subsection{Results}

\begin{figure}[tb]
 \begin{center}
  \includegraphics[width=80mm]{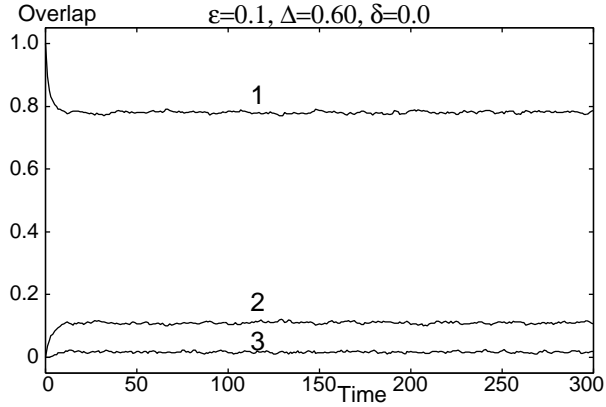} 
 \end{center}
 \caption{Time evolutions of overlaps $m^{1}_t$, $m^{2}_t$, and
 $m^{3}_t$ \rev{using a computer simulation} in the case without
 correlated noise, where $\varepsilon=0.1$ and $\Delta=0.6$.
 \label{fig:ovlp_no_eta} }
\end{figure}

The first case is without correlated noise $\eta^t$, that is,
$\delta=0.0$. Figure~\ref{fig:ovlp_no_eta} shows time evolutions of the
overlaps $m^{1}_t$, $m^{2}_t$, and $m^{3}_t$ using a computer
simulation, where the standard deviation (SD) of the independent noise
$\zeta_i^t$ is $\Delta=0.6$. The connecting parameter is
$\varepsilon=0.1$. Numbers in Figure~\ref{fig:ovlp_no_eta} denote
indexes of the memory patterns. We find that no transition occurs even
if large independent noise is induced. The model acts like the
autoassociative memory model through to the end.

\begin{figure}[tb]
 \begin{center}
  \includegraphics[width=75mm]{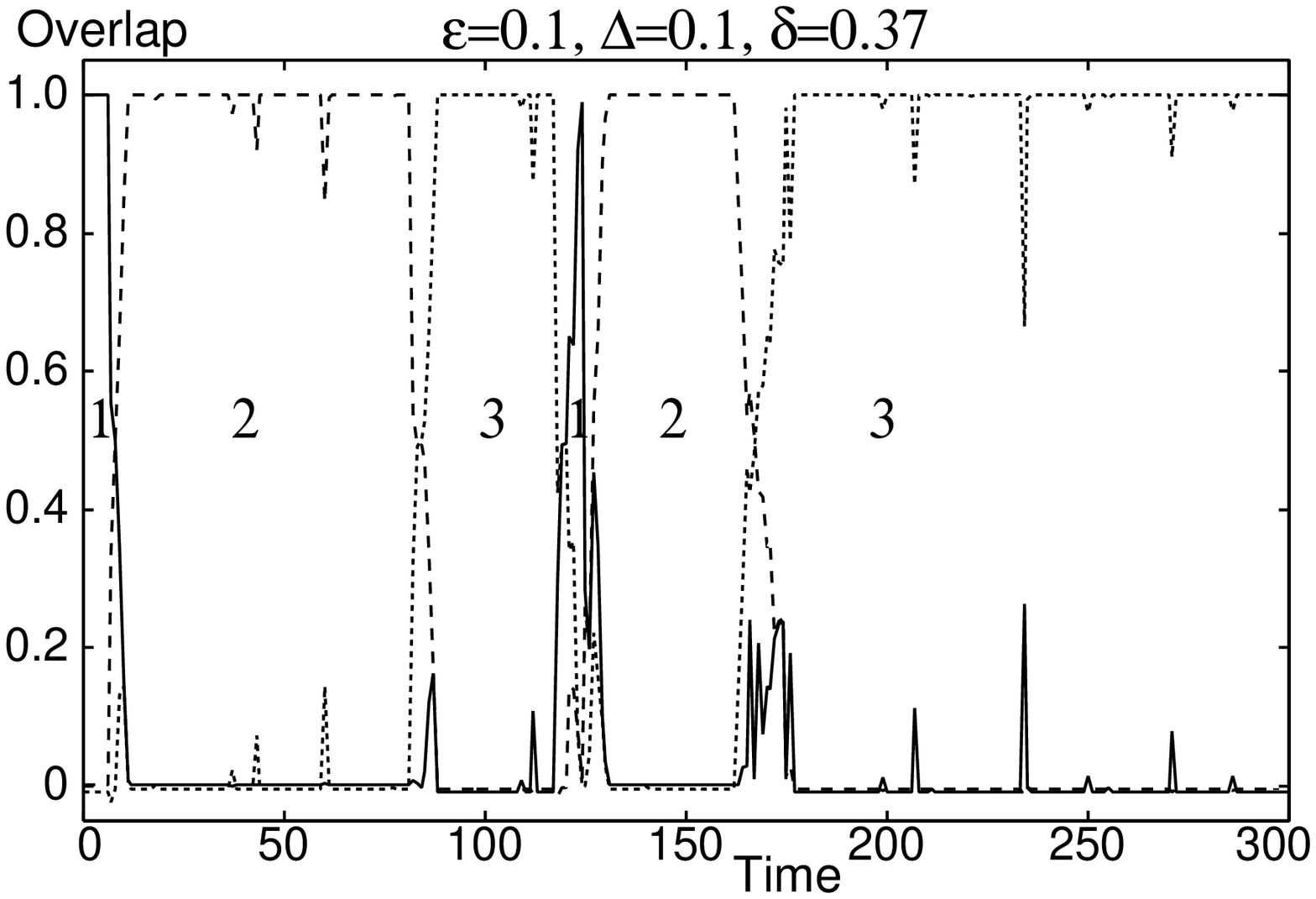}
  \includegraphics[width=75mm]{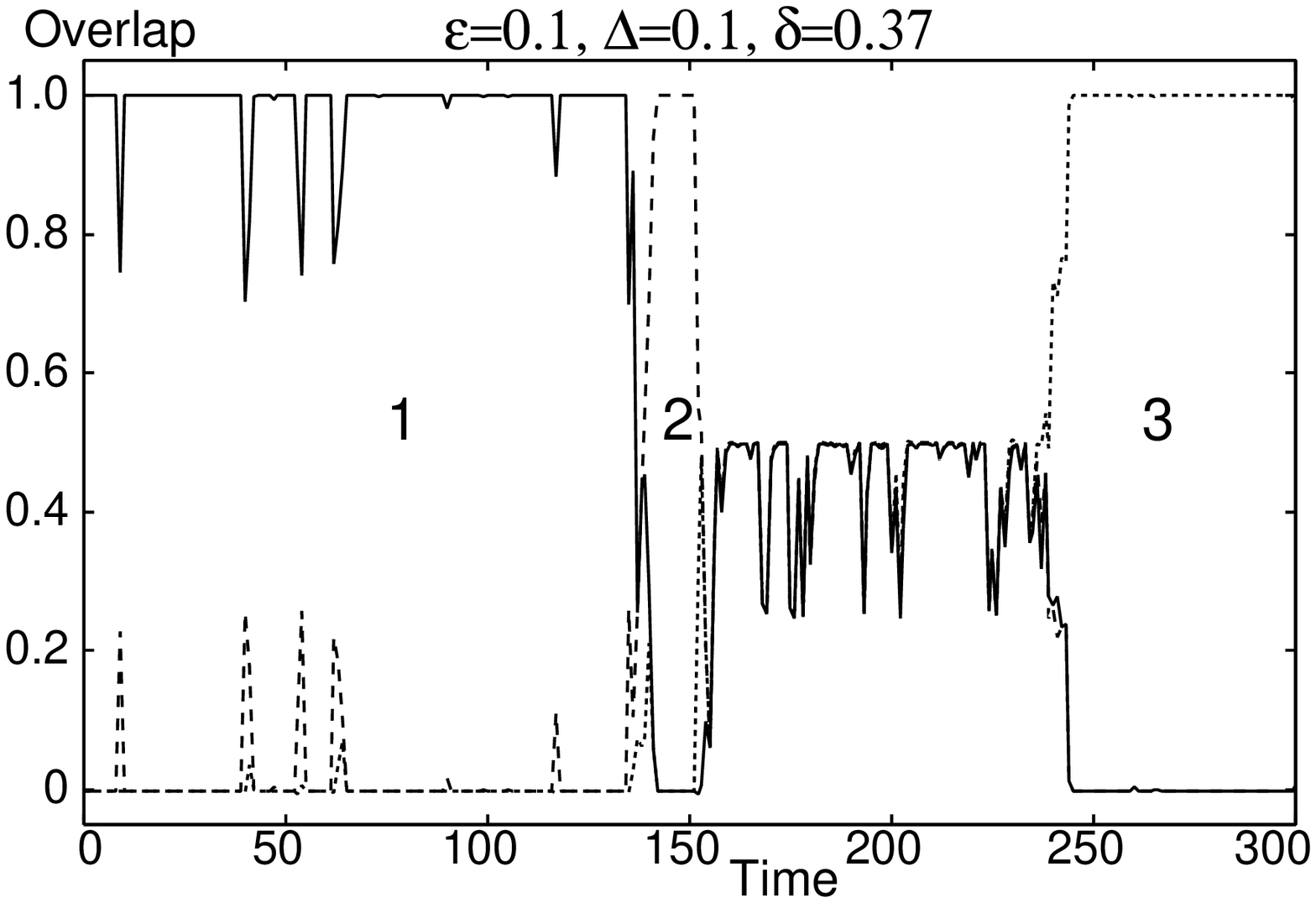}

  \hfill(a)\hfill\hfill(b)\hfill\mbox{}
 \end{center}
 \caption{Time evolutions of overlaps $m^{1}_t,m^{2}_t$, and $m^{3}_t$
 \rev{using computer simulations} in the case with correlated noise,
 where $\varepsilon=0.1, \Delta=0.6$, and $\delta=0.37$.
 \rev{(a) and (b) show results of different trials.}
 \label{fig:ovlp_eta} }
\end{figure}

Next, we consider the case with correlated noise
$\eta^t$. Figure~\ref{fig:ovlp_eta} shows time evolutions of the
overlaps $m^{1}_t$, $m^{2}_t$, and $m^{3}_t$ using computer simulations
($N=60,000$), where SDs of the independent and correlated noises are
$\Delta=0.1$ and $\delta=0.37$.  Figure~\ref{fig:ovlp_eta}(a) and (b)
show results of the different trials. Therefore, we can verify that
sample dependence exists and that the correlated noise invokes
stochastic transition.
The model acts like the sequential associative memory model.  In the
case of smaller correlated noise, transition probability to the next
attractor is much smaller. However, in the case of larger noise, the
state becomes much noisier, and we cannot identify what pattern is
retrieved. Therefore, the appropriately sized correlated noise should be
introduced in the model.

\begin{figure}[tb]
 \begin{center}
  \hfill\includegraphics[width=75mm]{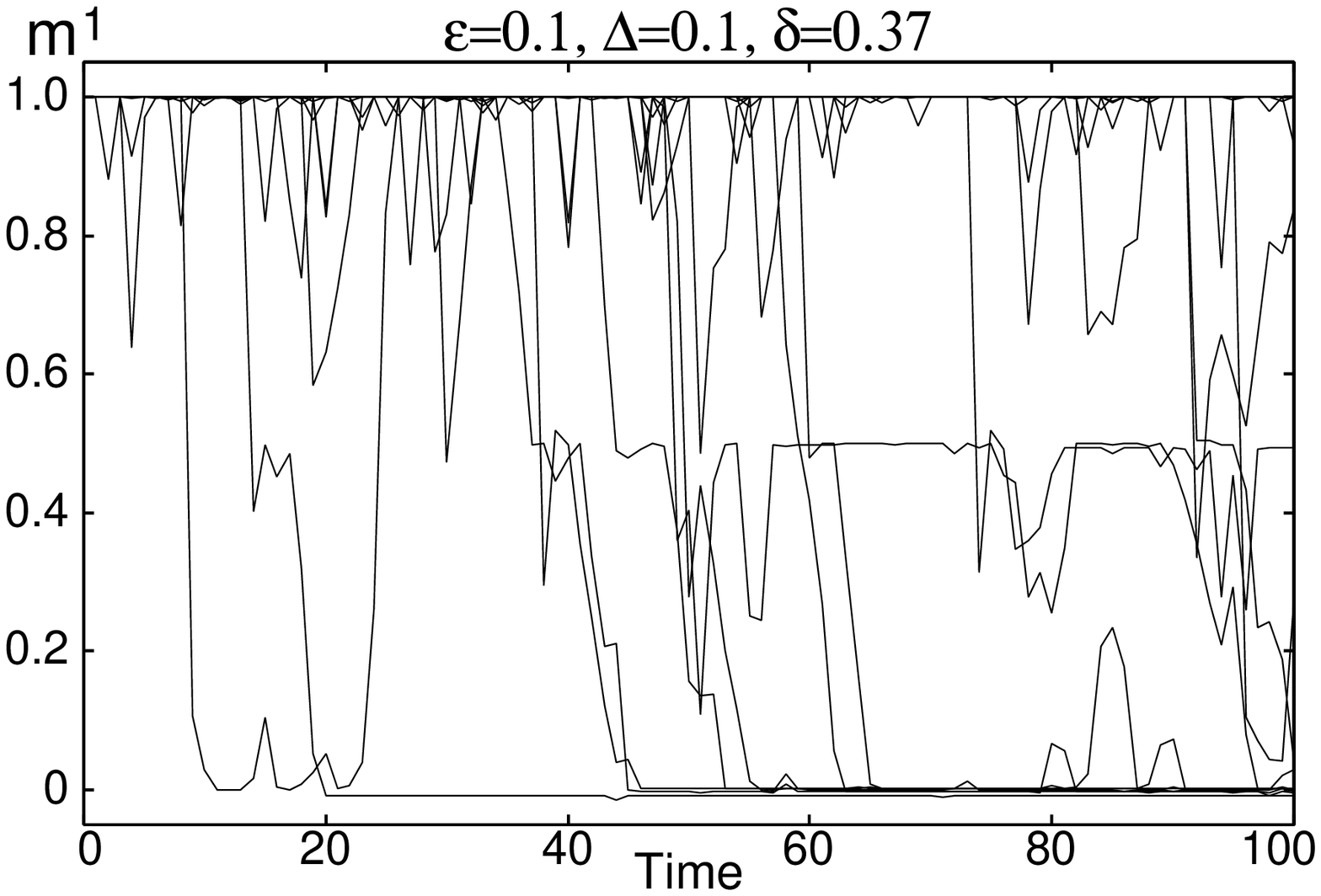}
  \hfill\includegraphics[width=75mm]{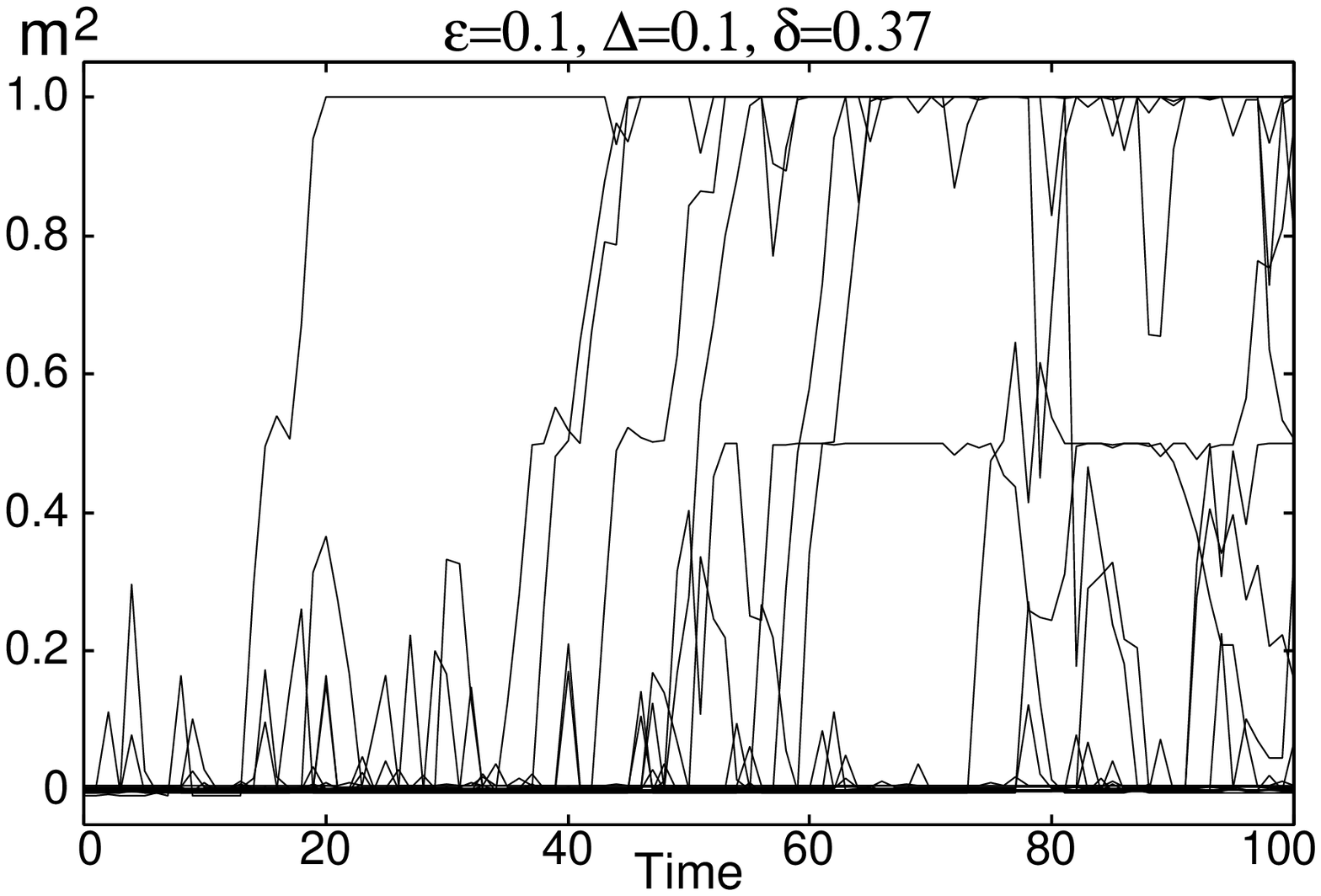}
  \hfill\mbox{}

  \hfill(a) Overlap $m^{1}_t$\hfill\hfill(b) Overlap $m^{2}_t$\hfill\mbox{}

  \includegraphics[width=75mm]{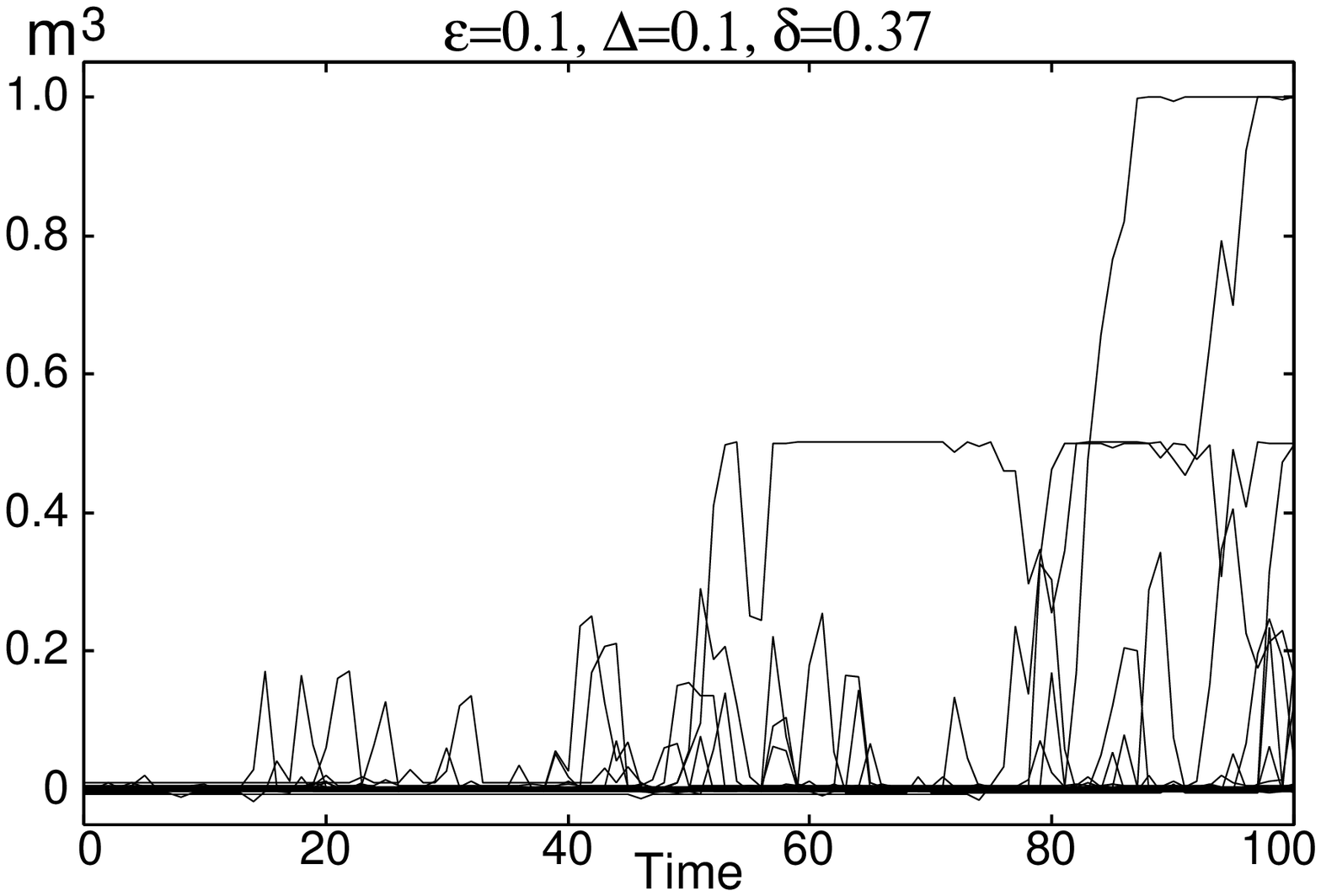}

  (c) Overlap $m^{3}_t$
 \end{center}
 \caption{\rev{20 samples of overlaps $m^{1}_t$, $m^{2}_t$, and $m^{3}_t$
 starting from $m_0=1.0$ using computer simulations, where
 $\varepsilon=0.1, \Delta=0.1$, and $\delta=0.37$.
 (a) shows results of overlap $m^{1}_t$, (b) for $m^{2}_t$, and (c) for
 $m^{3}_t$.}
 \label{fig:sample_depend} }
\end{figure}

\begin{figure}[htb]
 \begin{center}
  \hfill\includegraphics[width=75mm]{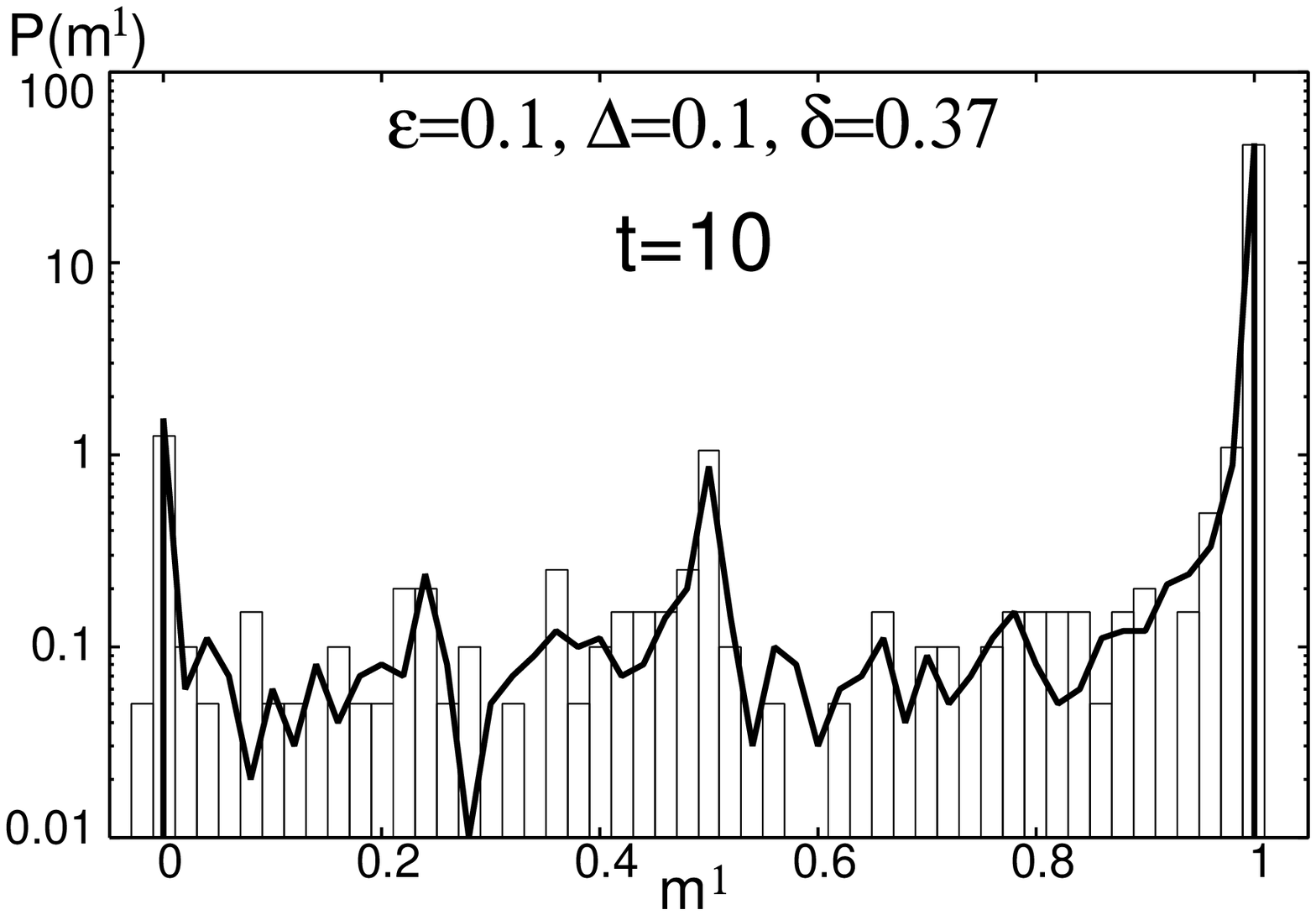}
  \hfill\includegraphics[width=75mm]{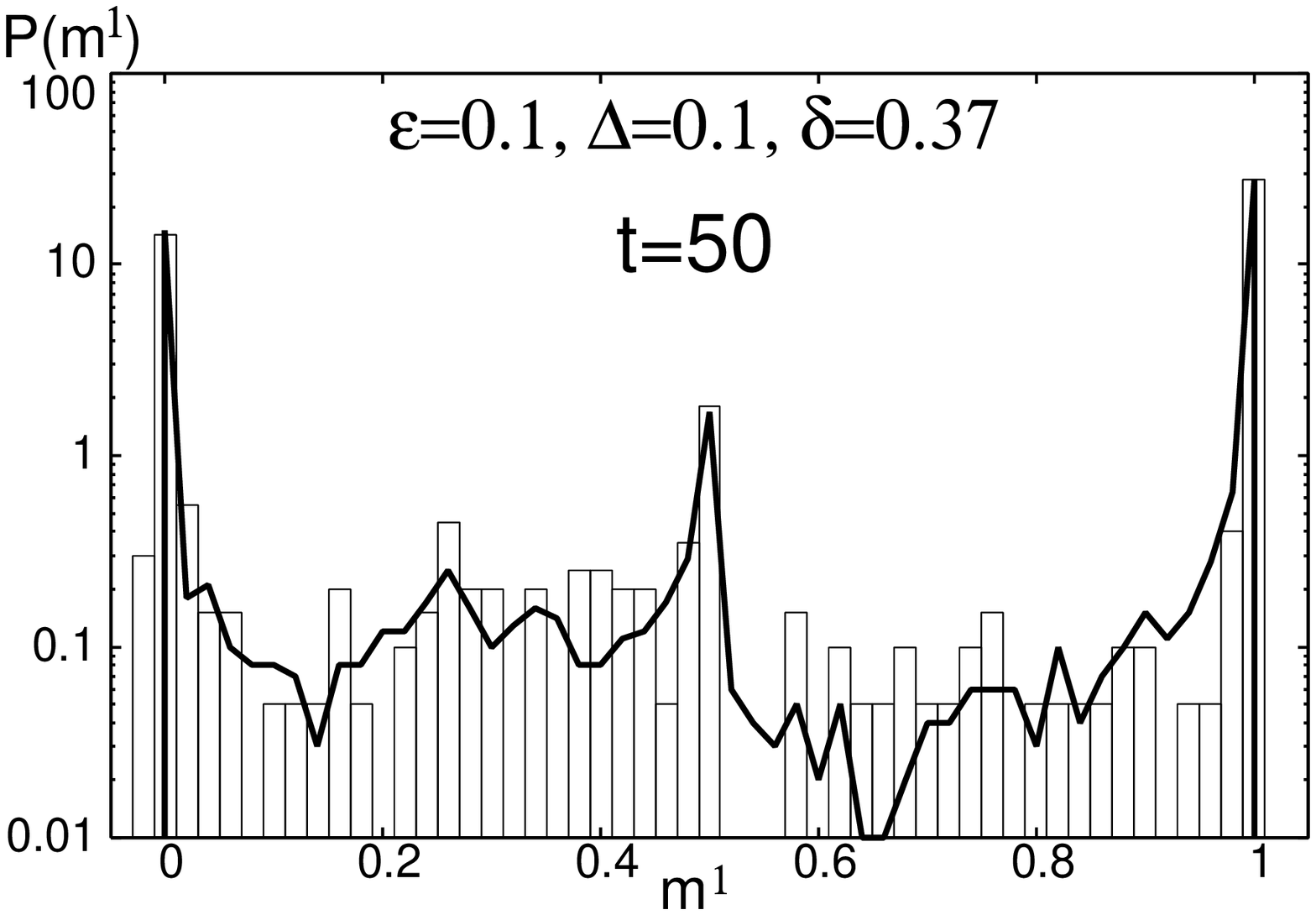}
  \hfill\mbox{}

  \hfill(a) $p\left(m^1_t\right)$ of pattern $\vec{\xi}^1$ at $t=10$
  \hfill(b) $p\left(m^1_t\right)$ of pattern $\vec{\xi}^1$ at $t=50$
  \hfill\mbox{}

  \hfill\includegraphics[width=75mm]{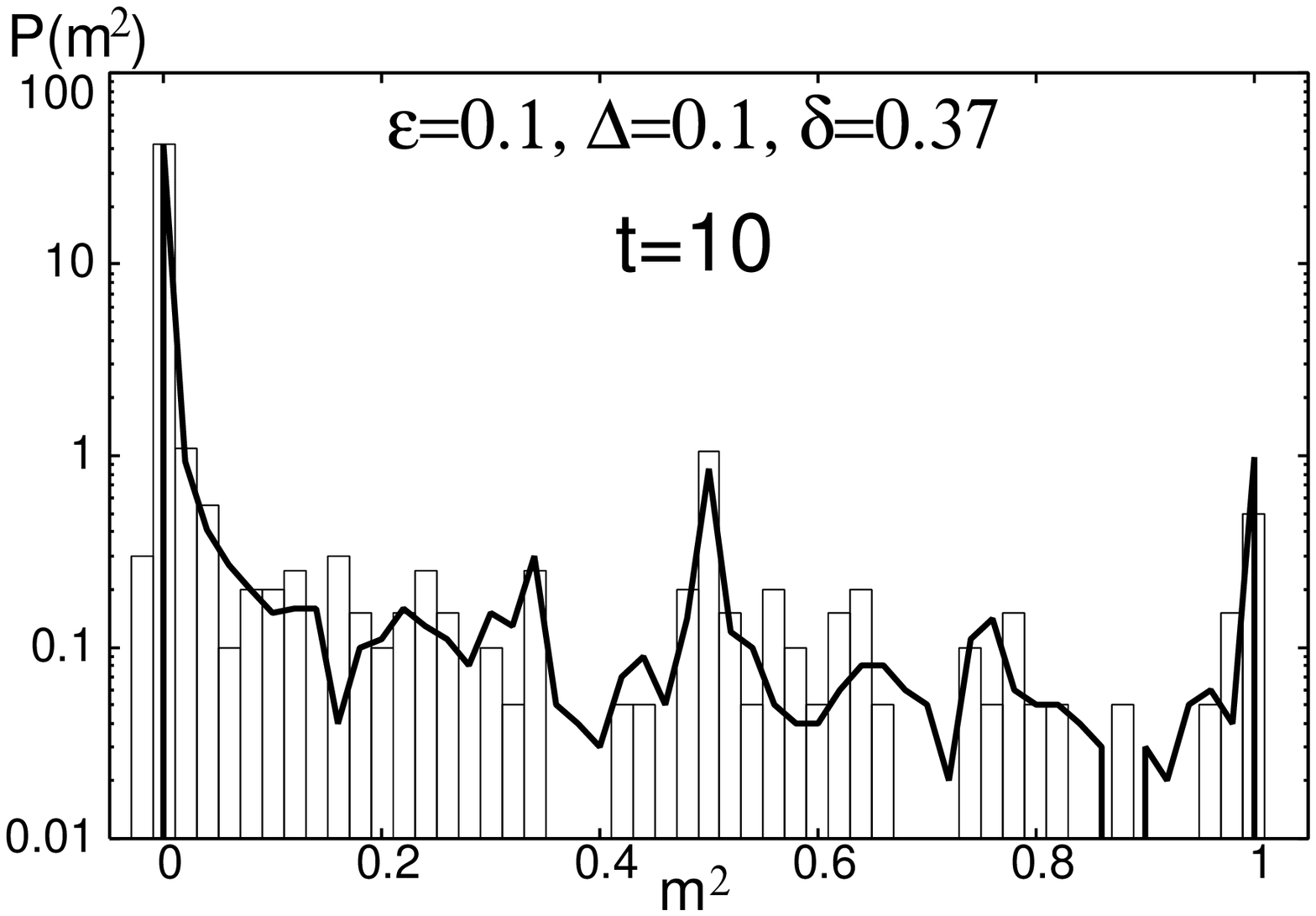}
  \hfill\includegraphics[width=75mm]{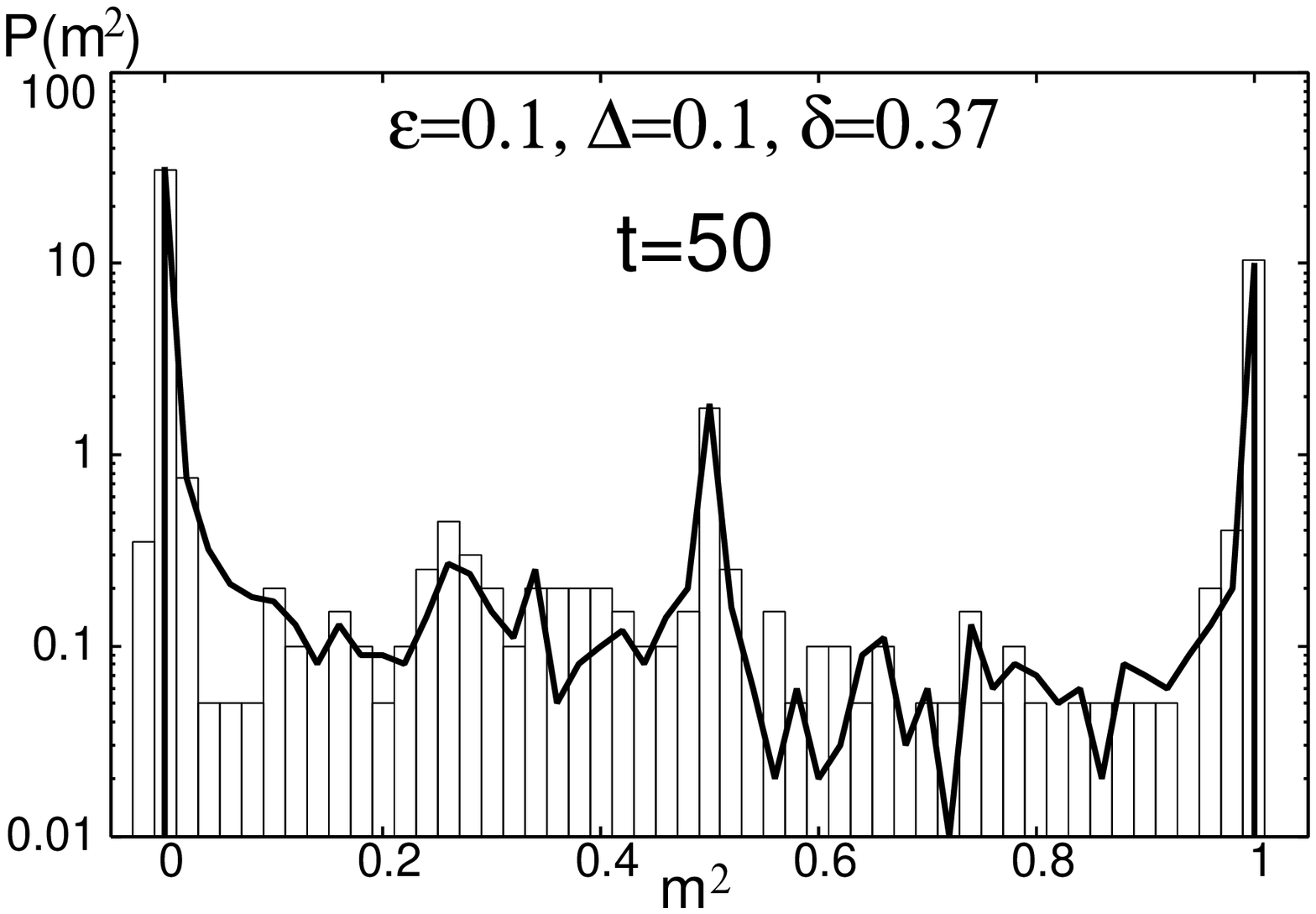}
  \hfill\mbox{}

  \hfill(c) $p\left(m^2_t\right)$ of pattern $\vec{\xi}^2$ at $t=10$
  \hfill(d) $p\left(m^2_t\right)$ of pattern $\vec{\xi}^2$ at $t=50$
  \hfill\mbox{}

  \hfill\includegraphics[width=75mm]{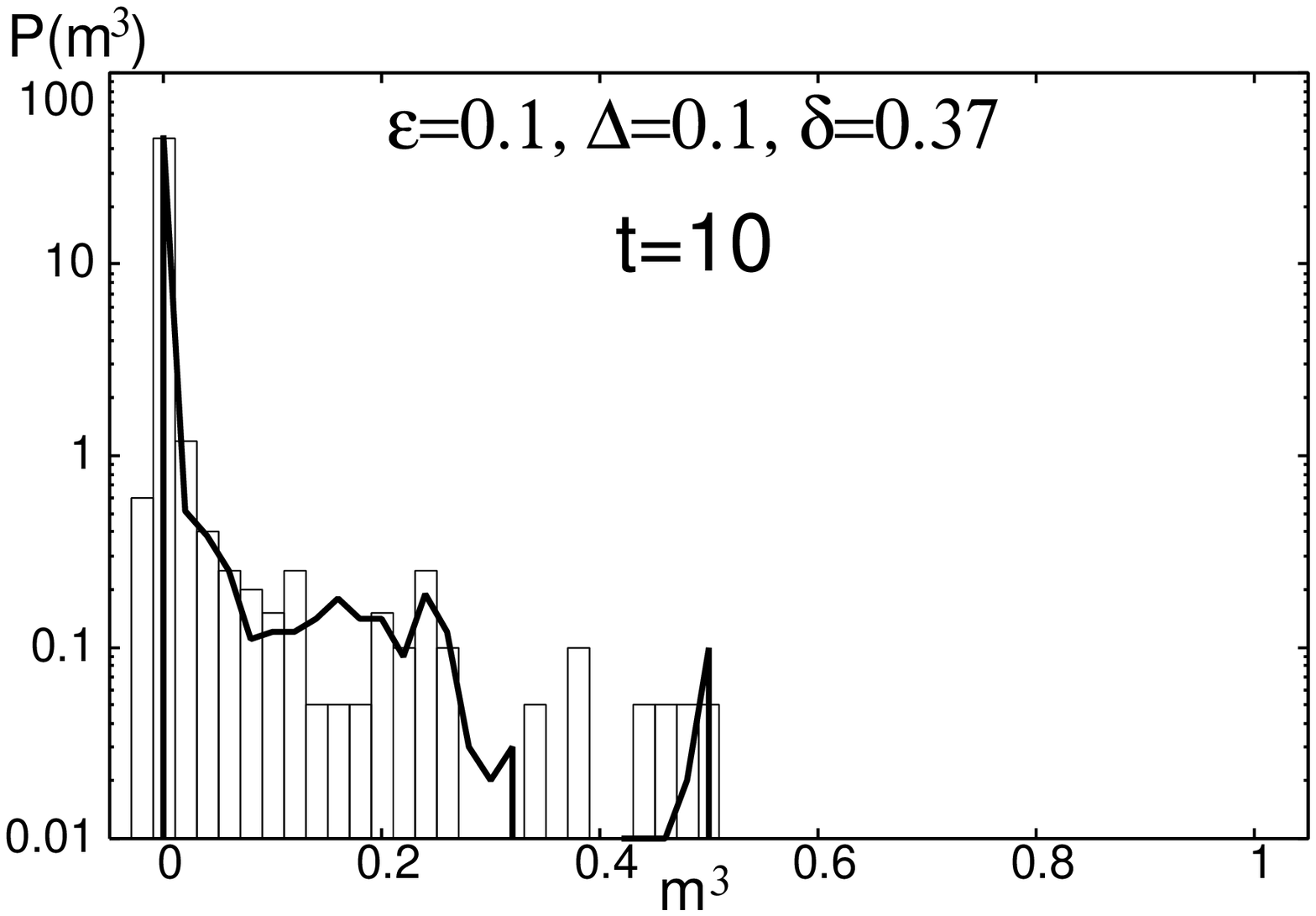}
  \hfill\includegraphics[width=75mm]{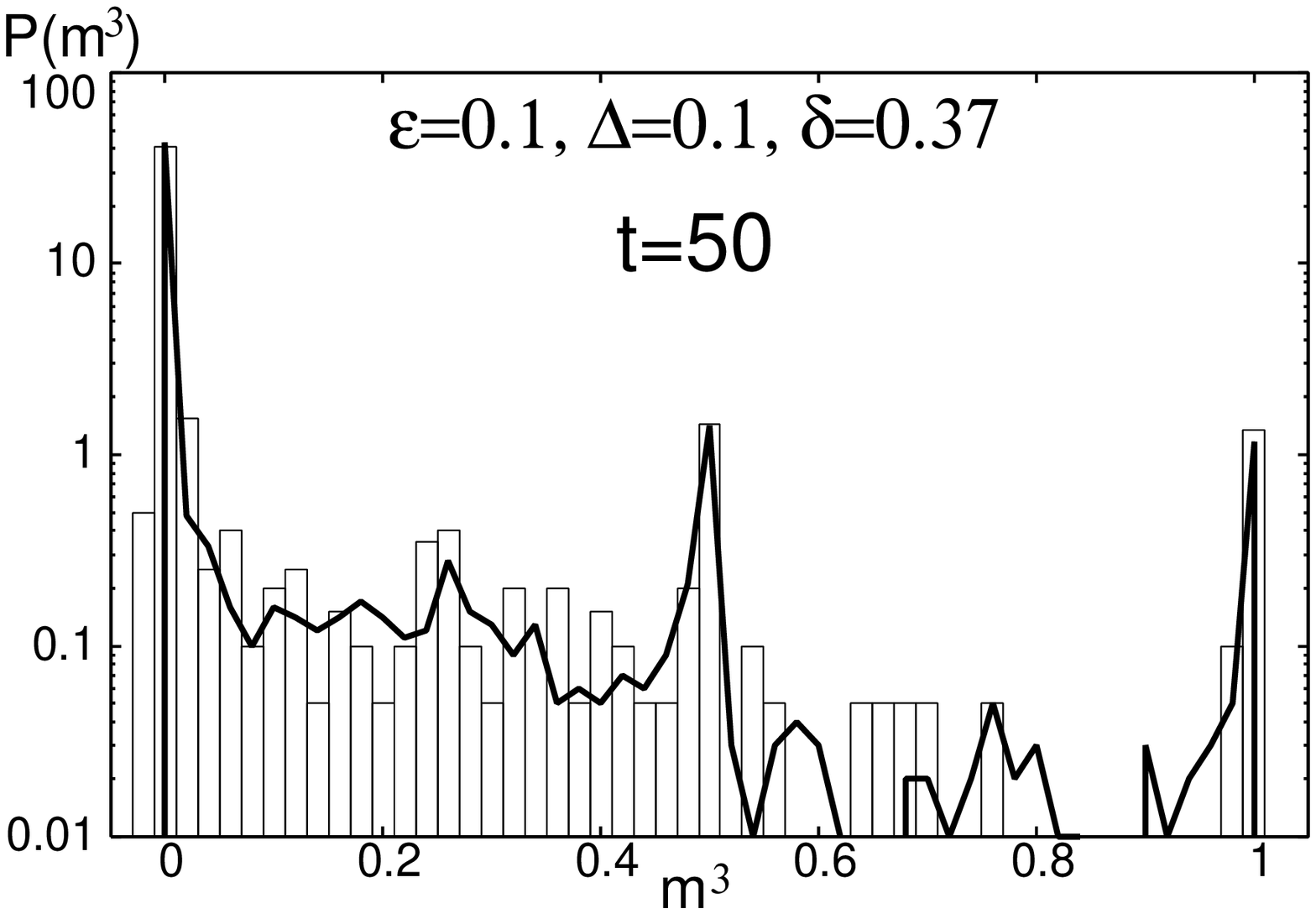}
  \hfill\mbox{}

  \hfill(e) $p\left(m^3_t\right)$ of pattern $\vec{\xi}^3$ at $t=10$
  \hfill(f) $p\left(m^3_t\right)$ of pattern $\vec{\xi}^3$ at $t=50$
  \hfill\mbox{}
 \end{center}
 \caption{Distribution of overlaps, $m^1_t$, $m^2_t$, and $m^3_t$ at
 time $t=10$, $50$,
 \rev{where $\varepsilon=0.1, \Delta=0.1$, and $\delta=0.37$.
 Abscissas denote overlap and ordinates denote marginal PDF on a
 logarithmic scale. Boxes denote histograms obtained using computer
 simulations, and lines denote theoretical results. 
 (a), (c), and (e) show results at $t=10$, and (b), (d), and (f) show
 results at $t=50$. }
 \label{fig:hist_seq}}
\end{figure}

Let us consider sample dependence. Figure~\ref{fig:sample_depend}
shows the time evolution of the overlaps $m^{1}_t$, $m^{2}_t$, and
$m^{3}_t$. The number of neurons is $N=60,000$. Each figure shows 20
samples. The initial overlap is $m_0=1.0$. When sample dependence
appears, we need to consider the distribution, not the overlaps, of the
samples.
From the joint probability density function
$p\left(m^{1}_t,m^{2}_t,m^{3}_t\right)$ in (\ref{eqn:PnextK}), we
evaluate the marginal probability density function of $m^{\mu}_t,
\mu=1,2,3$,
\begin{equation}
 p(m^{\mu}_{t})=\int \prod_{\nu\neq\mu}dm^{\nu}_t\; p\left(m^{1}_t,m^{2}_t,m^{3}_t\right).
  \label{eqn:Pmt}
\end{equation}
Figure~\ref{fig:hist_seq} shows the marginal PDF of the overlaps,
$p\left(m^1_t\right), p\left(m^2_t\right), p\left(m^3_t\right)$ at time
$t=10$, $50$ in Figure~\ref{fig:sample_depend}. Abscissas denote overlap
$m^{\mu}_t$, and ordinates denote marginal PDF $p\left(m^{\mu}_t\right)$
on a logarithmic scale. Boxes denote histograms of $1,000$ samples,
which are obtained using computer simulations ($N=60,000$), and lines
denote theoretical results. The theoretical results agree with the
computer simulations. The state at time $t=10$ distributes around the
memory state $\vec{\xi}^1$, since the overlap $m^1_{10}$ takes the
maximum in $m^1_{10}=1.0$. The state at time $t=50$ also distributes
around that of $\vec{\xi}^2$. As time passes, these PDFs approach a
uniform distribution asymptotically.

\begin{figure}[tb]
 \begin{center}
  \includegraphics[width=90mm]{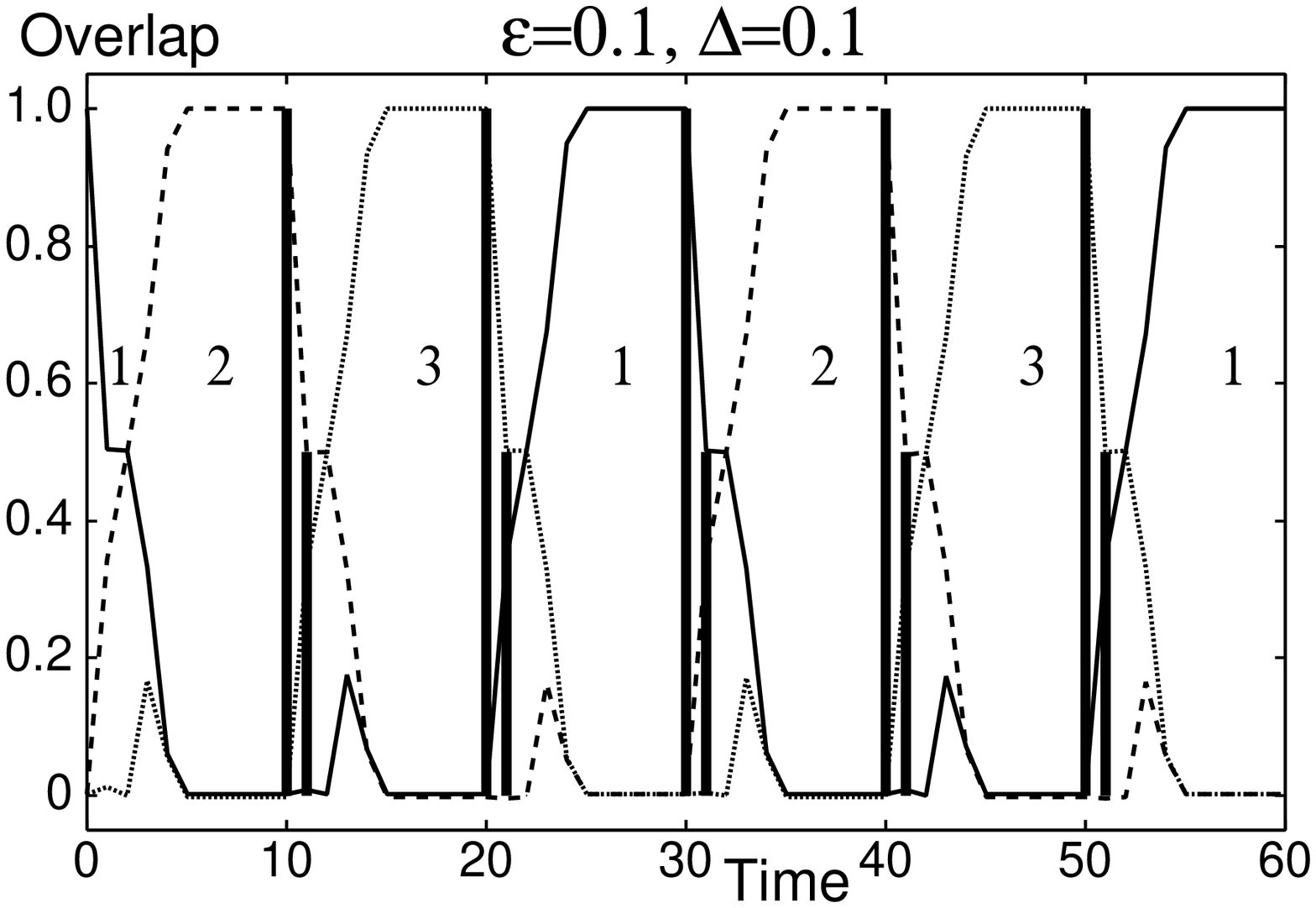}
 \end{center}
 \caption{Time evolutions of overlaps by particular common external input.
 \rev{Thick vertical lines denote the common external input. }
 }
 \label{fig:sample_const}
\end{figure}

We discussed the case when a common external input obeys the Gaussian
distribution. We can, however, choose alternative common external input.
The state transition seems to be caused by certain common external
input. We empirically found a particular common external
input. Figure~\ref{fig:sample_const} shows the state transition using
the following external input:
\begin{equation}
 \eta^t = \left\{\begin{array}{rcl}
           1 & , & $t \mbox{mod} 10 = 0$ \\
                  0.5 & , & $t \mbox{mod} 10 = 1$ \\
                  0 & , & otherwise \\
                 \end{array}\right. .
\end{equation}
Thick vertical lines denote the particular common external input
$\eta^t$.  For instance, if a sequence of the external inputs is
$\eta^{10}=1$ and $\eta^{11}=0.5$, then the state transition
$\vec{\xi}^2\to\vec{\xi}^3$ can occur the next time. When this is
repeated, the memory patterns are retrieved sequentially.  In this
manner, the common external input to the neurons can control the
transitions.

\section{Associative memory model with hierarchically correlated patterns}
\label{sec:hierar}

\subsection{Model}

Now we discuss the associative memory model with hierarchically
correlated patterns. Parent patterns $\xi^{\mu}_i, \mu=1,2\cdots,p$ are
generated at random, and then child patterns $\xi^{\mu,\nu}_i,
\nu=1,2,\cdots,k$ are generated from the parent patterns $\xi^{\mu}_i$
with similarity $r$, that is, they are given by
\begin{eqnarray}
 P\left[\xi^{\mu}_i=\pm1\right] &=& \frac{1}{2}, \\
 P\left[\xi^{\mu,\nu}_i=\pm1\right] &=& \frac{1+r\xi^{\mu}_i}{2}.
\end{eqnarray}
The child patterns having different parents are independent of each
other, but those having the same parent are correlated. In the limit of
$N\to\infty$, the correlations become
\begin{eqnarray}
 E\left[\xi^{\mu,\nu}_i\xi^{\tilde{\mu},\tilde{\nu}}_i\right]
  &=& 0 \;\;,\;\; \mu \neq \tilde{\mu} \; , \\
 E\left[\xi^{\mu,\nu}_i\xi^{\mu,\tilde{\nu}}_i\right]
  &=& \left\{\begin{array}{rr}
   1 \;,& \nu = \tilde{\nu} \\
       r^2 \;, & \nu \neq \tilde{\nu} \\
      \end{array} \right. \; .
\end{eqnarray}
The synaptic connection stores only the child patterns $\xi^{\mu,\nu}_i$
and is given by
\begin{equation}
 J_{ij}=\frac1N\sum_{\mu=1}^{p}\sum_{\nu=1}^{k}\xi^{\mu,\nu}_i\xi^{\mu,\nu}_j.
  \label{eqn:Jijc}
\end{equation}
Mixture states of the child patterns also become attractors even if only
the child patterns are stored \cite{AmitGutfreund1985}.

To make the argument simple, we assume that the numbers of the parent
and child patterns, $p$ and $k$, are finite, especially when $p=1$ and
$k=3$. When the similarity is $r=0$, that is, the child patterns are
independent of each other, the mixture states become unstable faster than
the memory states with an increase in temperature \cite{Hertz1991}.
When $r>0$, the mixture states become stabler.  We define the
overlap by the direction cosine between the state of neurons,
$\vec{x}^t$, at time $t$ and the child pattern $\vec{\xi}^{1,\mu}$,
\begin{equation}
 m^{\mu}_t = \frac{1}{N}\sum_{i=1}^N \xi_i^{1,\mu} x_i^t .
  \label{eqn:mtc}
\end{equation}
From Equations (\ref{eqn:dynamics}), (\ref{eqn:Jijc}), and (\ref{eqn:mtc}),
the state $x_i^{t+1}$ becomes
\begin{eqnarray}
 x_i^{t+1} &=& F \left(\sum_{\kappa=1}^{3} \xi^{1,\kappa}_i m^{\kappa}_t
		  +\zeta_i^t+\eta^t \right)  .
 \label{eqn:x_xic}
\end{eqnarray}

\subsection{Theory}

As in the case of the previous model, let us derive the PDF of this
model. In the case without correlated noise $\eta^t$, when
$N\to\infty$, the overlap $m^{\mu}_{t+1}$ can be 
\begin{eqnarray}
 m^{\mu}_{t+1}
 &=& \left< \int  D_z \xi^{1,\mu}
      F\left(\sum_{\kappa=1}^{3}\xi^{1,\kappa}m^{\kappa}_t+\Delta z\right)
       \right>_{\xi} , \\
 &=& \left< \xi^{1,\mu}
      \erf\left(\frac{\sum_{\kappa=1}^{3}\xi^{1,\kappa}m^{\kappa}_t}
           {\sqrt{2}\Delta}\right)\right>_{\xi} ,
      \label{eqn:merfc}
\end{eqnarray}
where $\left<\cdot\right>_{\xi}$ denotes the average over not only the child
patterns $\vec{\xi}^{1,\kappa}$, but also their parent pattern
$\vec{\xi}^{1}$.

In the case that $\eta^t$ would be known at given time $t$ and
$N\to\infty$, then $m^{\mu}_{t+1}$ could be given as the function of
$m^{1}_t$, $m^{2}_t$, $m^{3}_t$, and $\eta^t$,
\begin{eqnarray}
 m^{\mu}_{t+1}(m^{1}_t,m^{2}_t,m^{3}_t,\eta^t)
 &=& \left<\int \!\! D_{z}
      \xi^{1,\mu} F\left(\sum_{\kappa=1}^3\xi^{1,\kappa}m_t^{\kappa}
                    +\Delta z+\eta^t\right)\right>_{\xi},  \\
 &=& \left< \xi^{1,\mu}
      \erf\left(\frac{\sum_{\kappa=1}^{3}\xi^{1,\kappa}m^{\kappa}_t+\eta^t}
           {\sqrt{2}\Delta}\right)\right>_{\xi} .
\end{eqnarray}
Substituting Equations (\ref{eqn:PnextK}) and (\ref{eqn:Kernel}), we can obtain
the recurrence relation form of the PDF for this model.

\subsection{Function of External Noise}

\subsubsection{Without correlated noise ($\eta^t=0$)}

\begin{figure}[tb]
 \begin{center}
  \hfill\includegraphics[width=70mm]{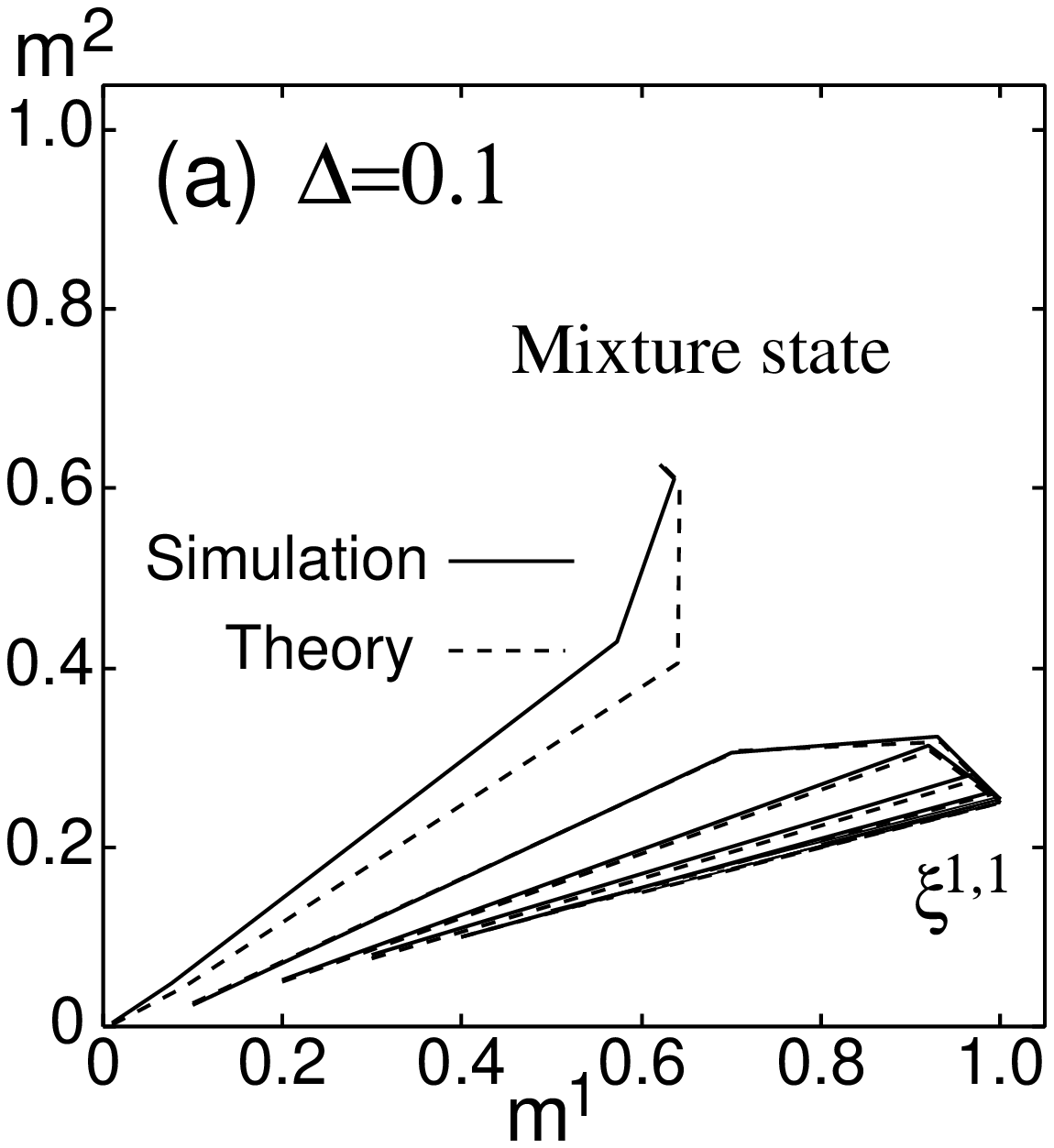}
  \hfill\includegraphics[width=70mm]{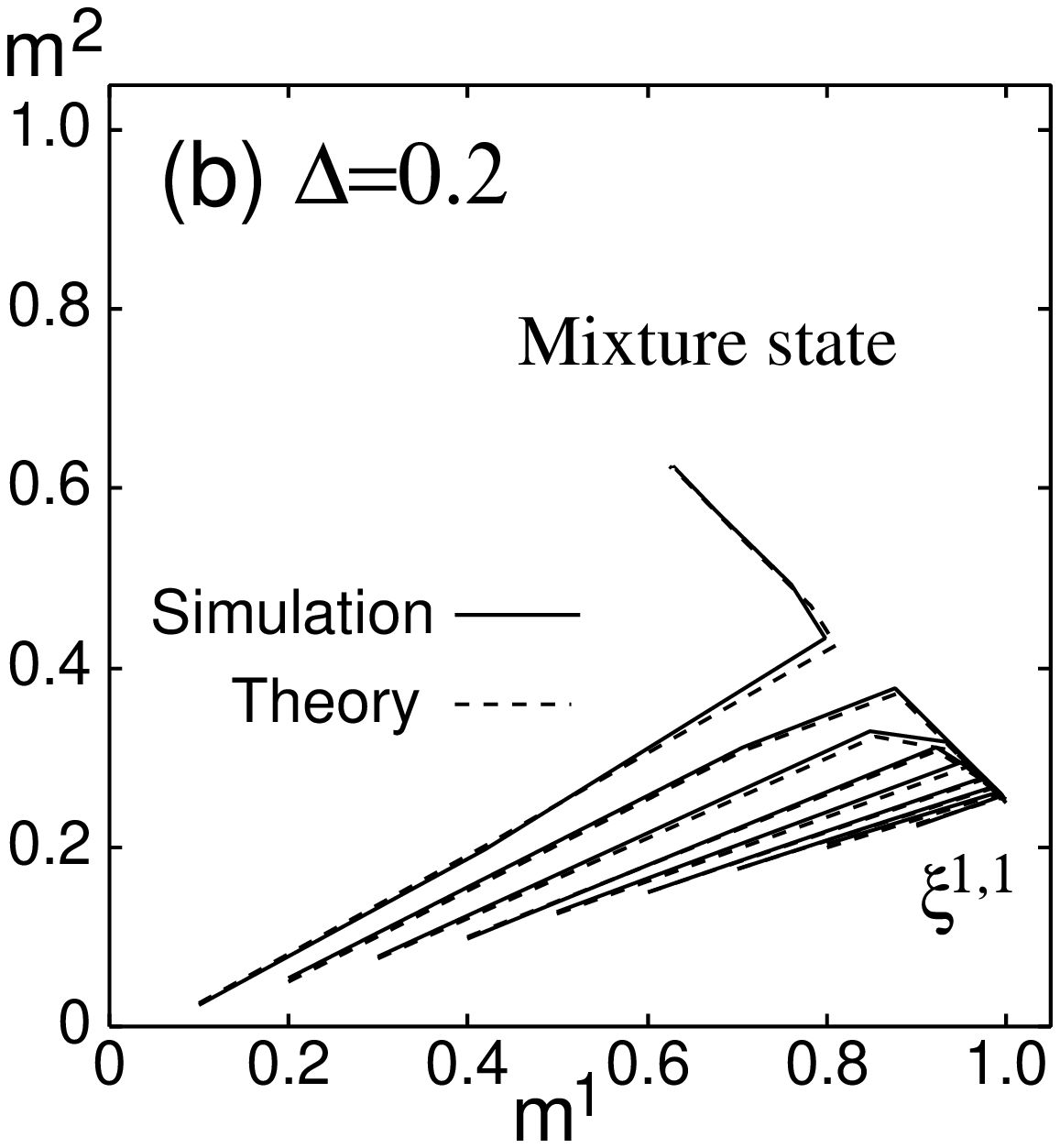}
  \hfill\mbox{}

  \hfill\includegraphics[width=70mm]{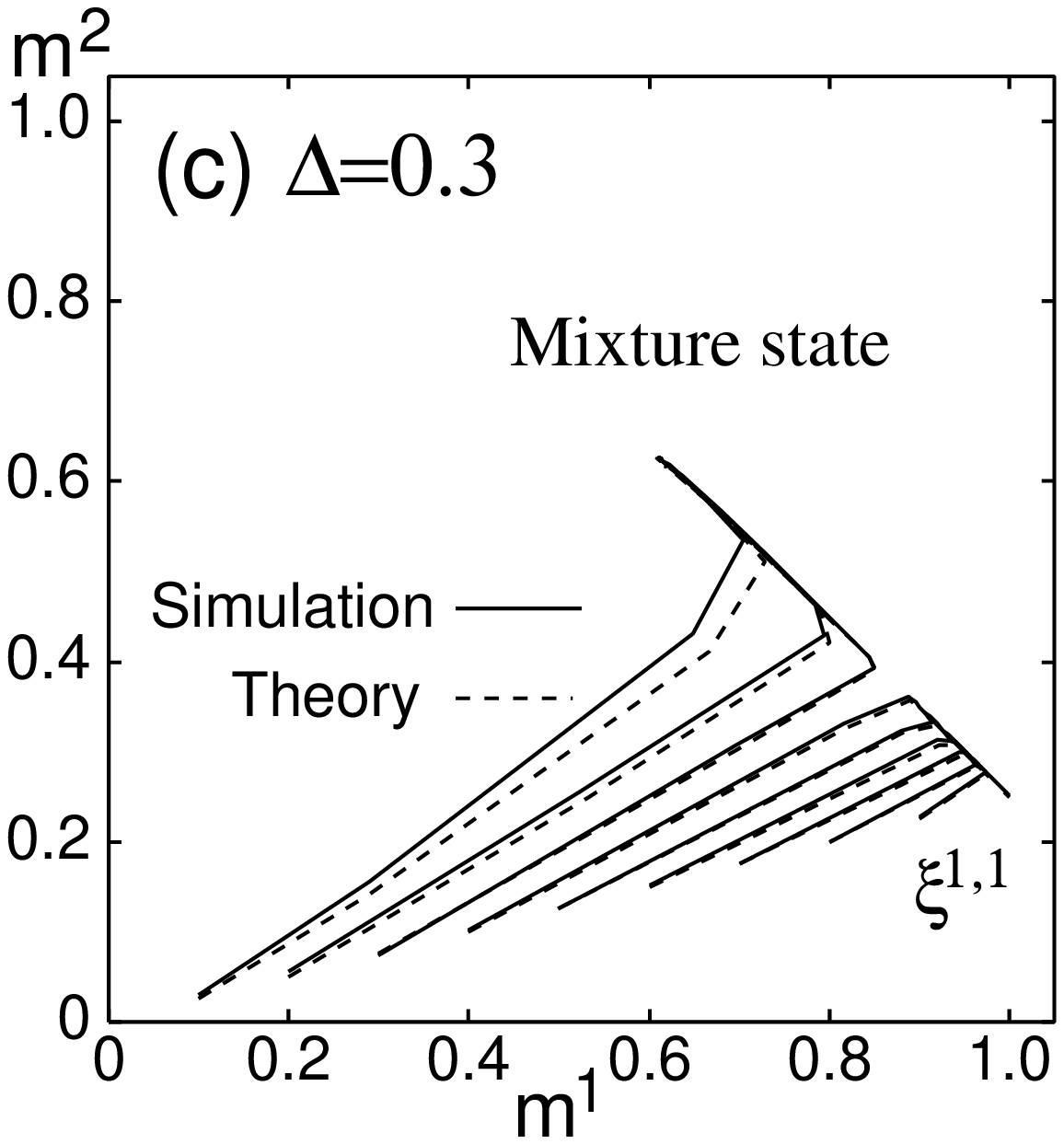}
  \hfill\includegraphics[width=70mm]{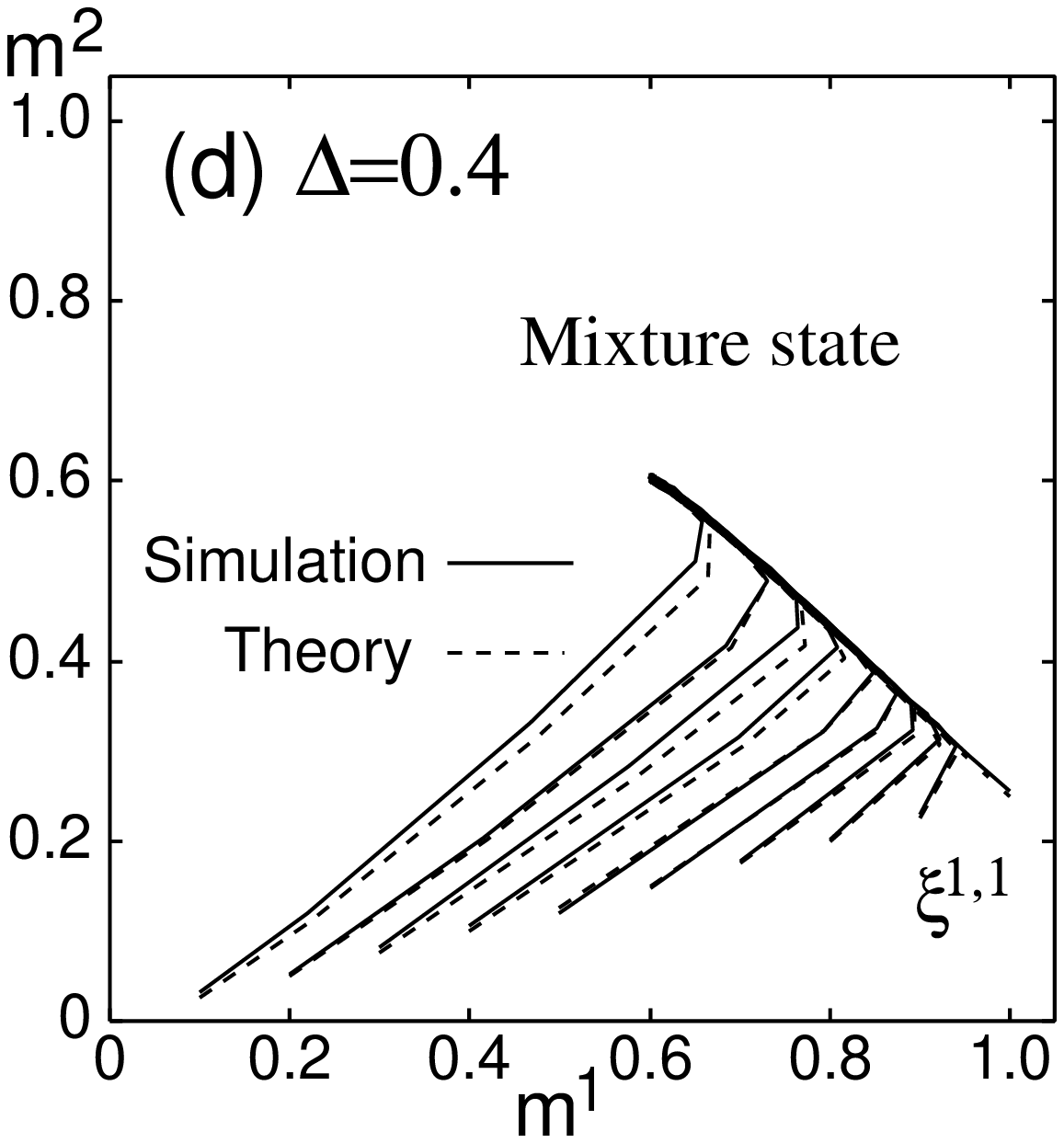}
  \hfill\mbox{}
 \end{center}
 \caption{Retrieval processes on $(m^1_t,m^2_t)$-plane, where $r=0.5$ and
 (a)~$\Delta=0.1$, (b)~$\Delta=0.2$, (c)~$\Delta=0.3$, and (d)~$\Delta=0.4$.
 \label{fig:m12Z0.2}}
\end{figure}

We show dynamic behaviors without correlated noise $\eta^t$. To verify
retrieval of a memory state of a child pattern, we evaluated retrieval
processes for various initial overlaps. Figure~\ref{fig:m12Z0.2} shows
the retrieval processes for the initial overlaps
$m_0=0.1,0.2,\cdots,1.0$. The abscissas and ordinates denote the
overlaps $m^1_t$ and $m^2_t$. Solid and broken lines denote results
obtained using computer simulations ($N=60,000$) and Equation
(\ref{eqn:merf}).  When $\Delta\leq0.3$, the state is attracted to the
attractor of $\vec{\xi}^{1,1}$ for a significantly large initial
overlap. When $\Delta=0.4$ or has a small initial overlap, the state
goes to mixture state, not the memory state. We find that independent
noise cannot invoke a transition to the mixture state once the state is
attracted to the memory state.

\subsubsection{With correlated noise ($\eta^t\neq0$)}

\begin{figure}[tb]
 \begin{center}
  \hfill\includegraphics[width=75mm]{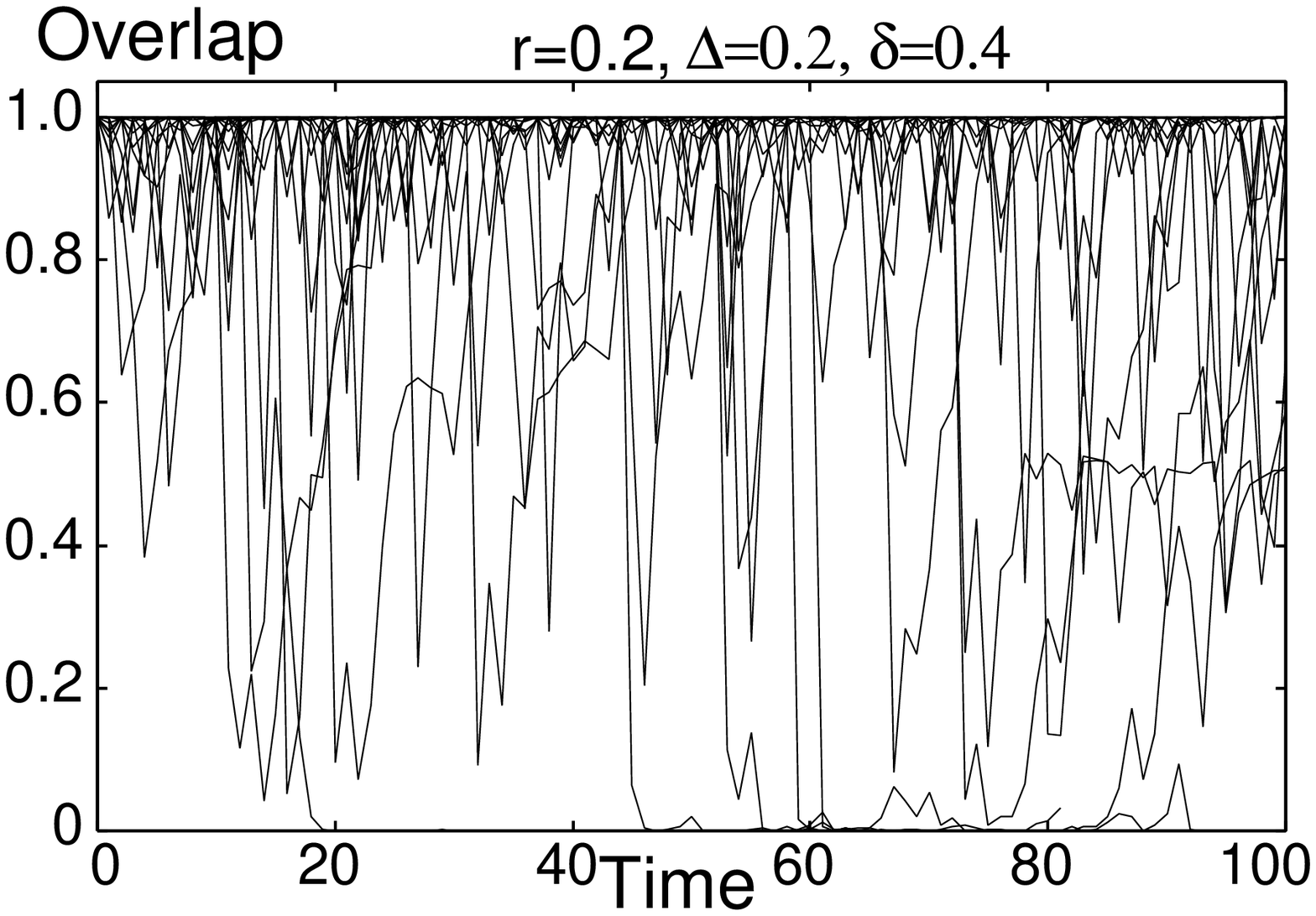}
  \hfill\includegraphics[width=75mm]{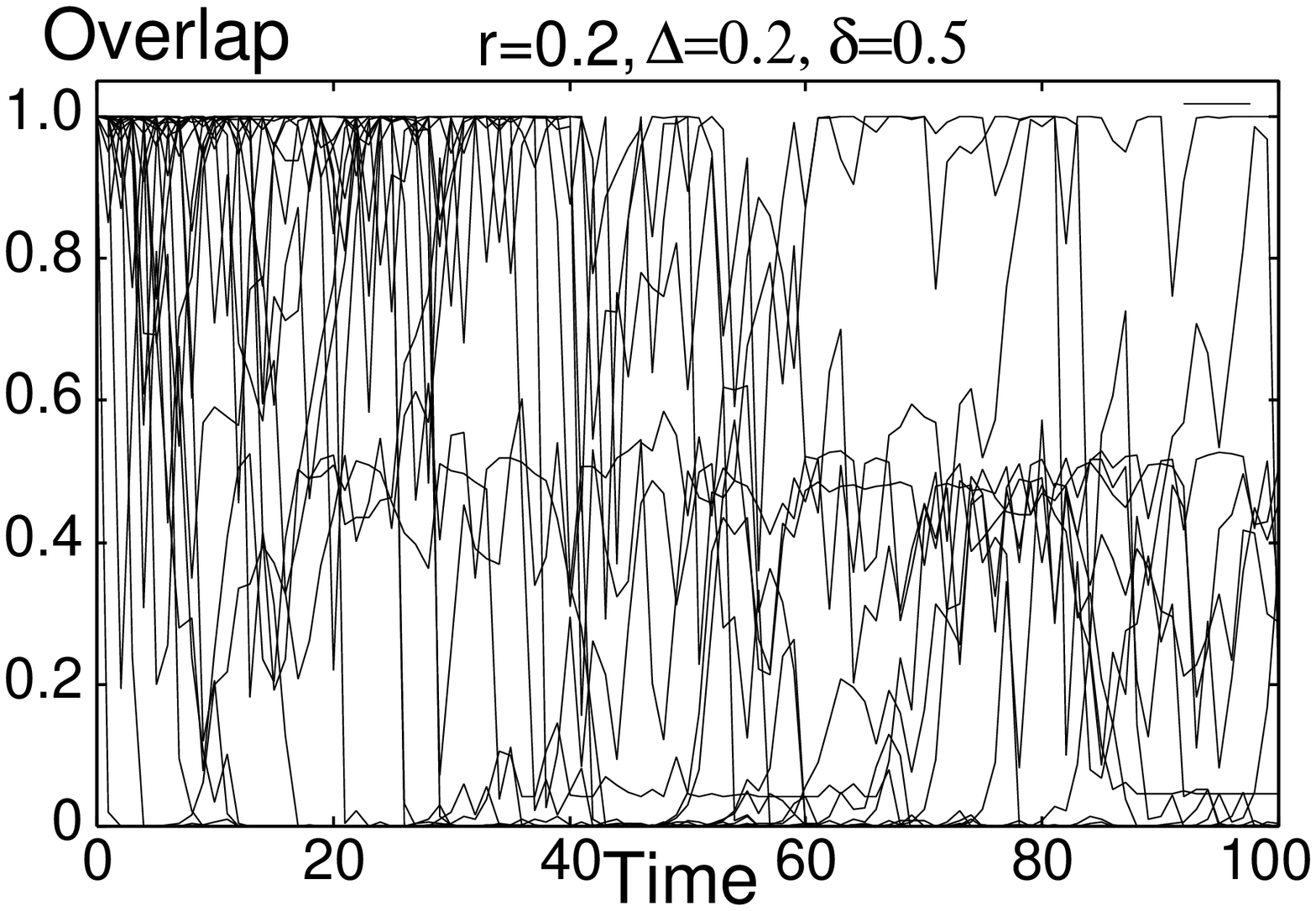}
  \hfill\mbox{}
  
  \hfill(a)~$\delta=0.4$ \hfill\hfill(b)~$\delta=0.5$ \hfill\mbox{}
 \end{center}
 \caption{Time evolution of overlap with correlated noise 
 \rev{for (a) $\delta=0.4$, and (b) $\delta=0.5$,
 where $r=0.2$ and $\Delta=0.2$. 
 The results are obtained from 20 samples using computer simulations.
 }
 \label{fig:ovlp_d0.4-0.5}}
\end{figure}

We show dynamic behavior with correlated noise $\eta^t$ using computer
simulations. In the case of $\delta>0$, sample dependence appears and
the network retrieves either memory or non-memory state even if
retrieval from the same initial state.
To examine functional roles of the correlated noise, we choose a
sufficiently small independent noise, $\Delta=0.2$.
Figure~\ref{fig:ovlp_d0.4-0.5} shows time evolutions of the overlap with
correlated noise. The results are obtained from 20 samples using computer
simulations ($N=60,000$), where $\delta=0.4$ or $0.5$ and $r=0.2$. 
Sample dependence occurs. When $\delta=0.4$, the state is almost
wrapped around the memory state, and when $\delta=0.5$, it moves
around the mixture state in several samples. Correlated noise can
cause a transition from the memory to the mixture state
stochastically.

\begin{figure}[tb]
 \begin{center}
  \includegraphics[width=80mm]{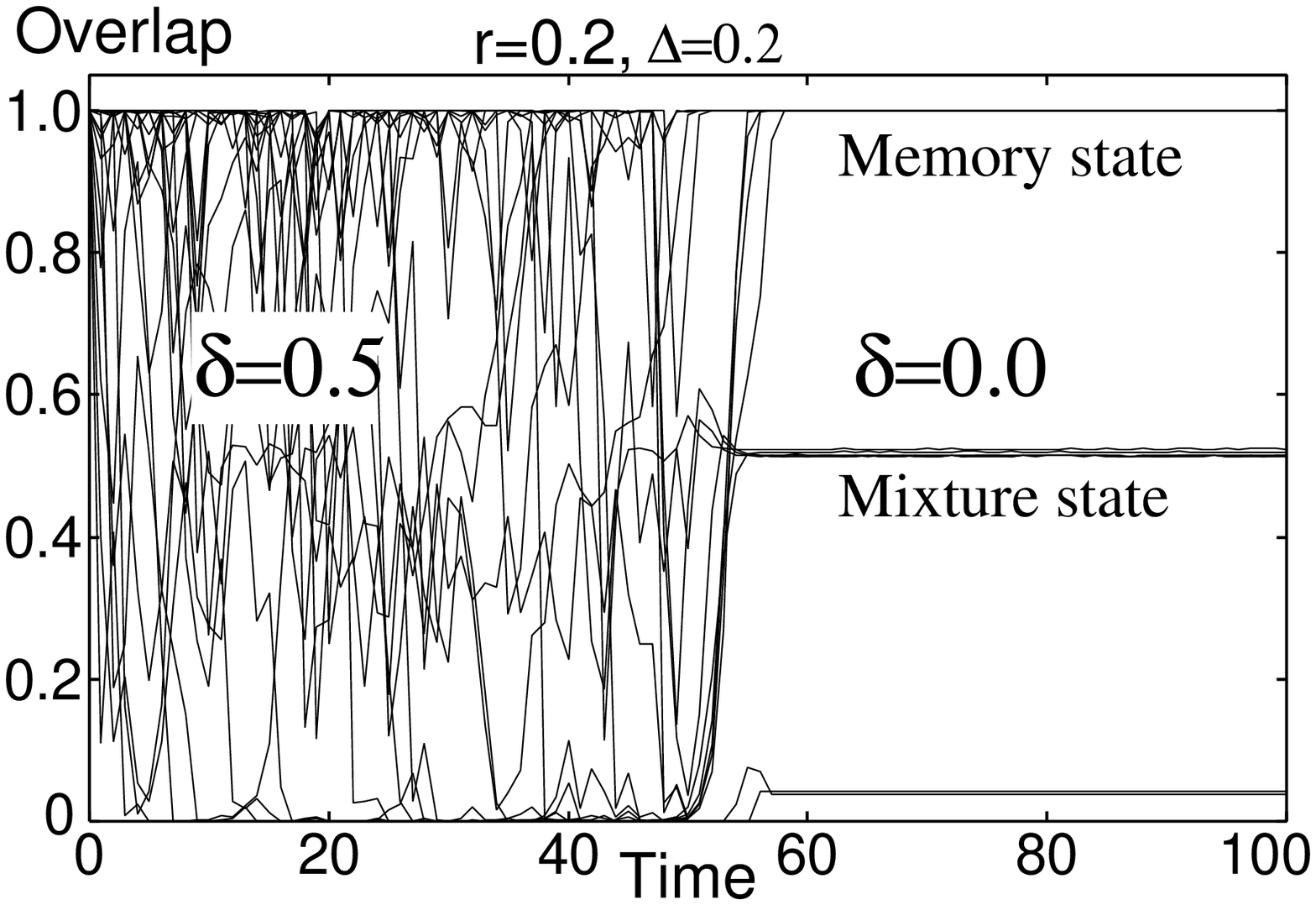}
 \end{center}
 \caption{\rev{Convergence of overlaps around the mixture state.}
 Correlated noise is induced until $t<50$. 
 \label{fig:ovlp_d0.5-0}}
\end{figure}

\begin{figure}[tb]
 \begin{center}
  \hfill\includegraphics[width=75mm]{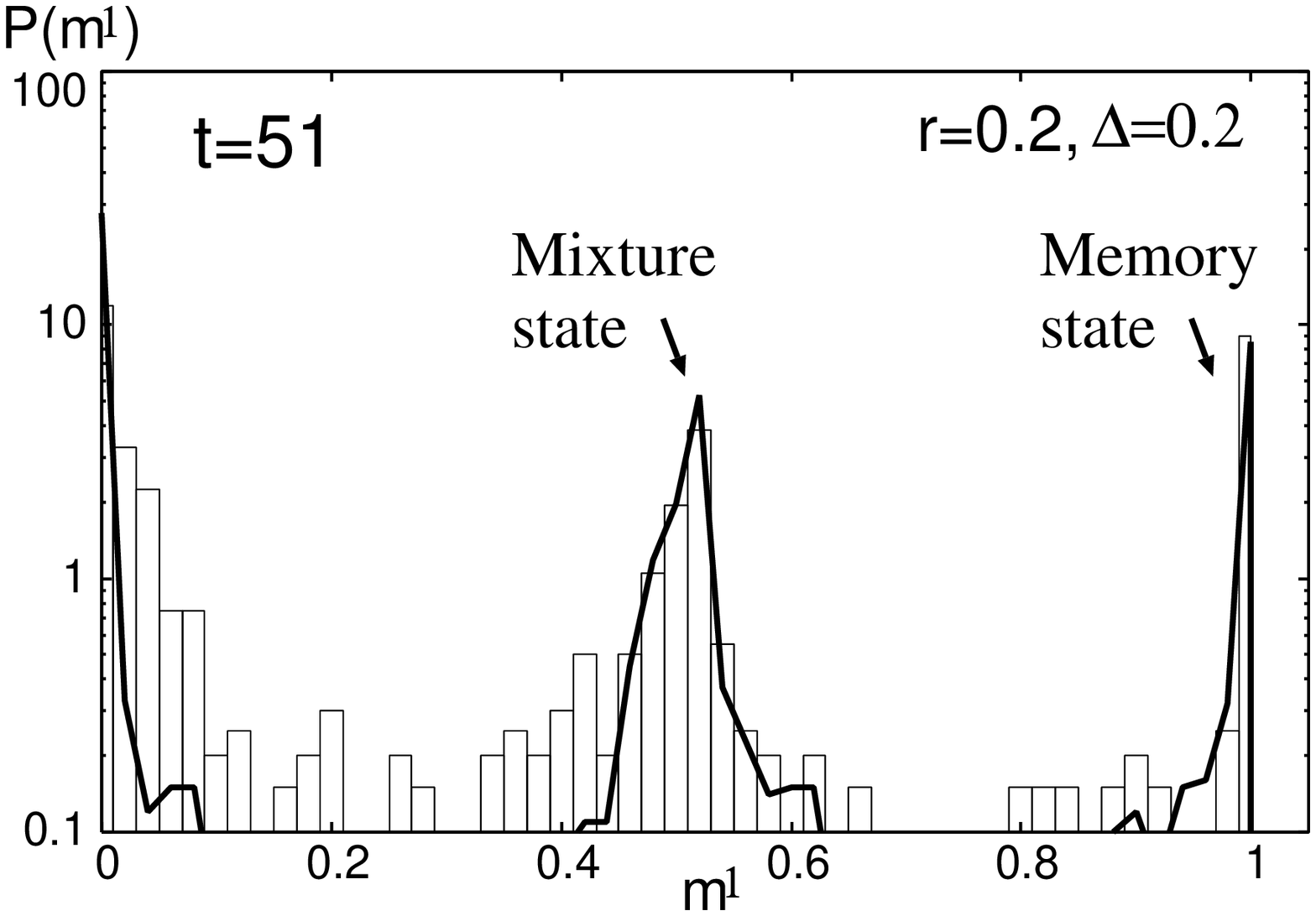}
  \hfill\includegraphics[width=75mm]{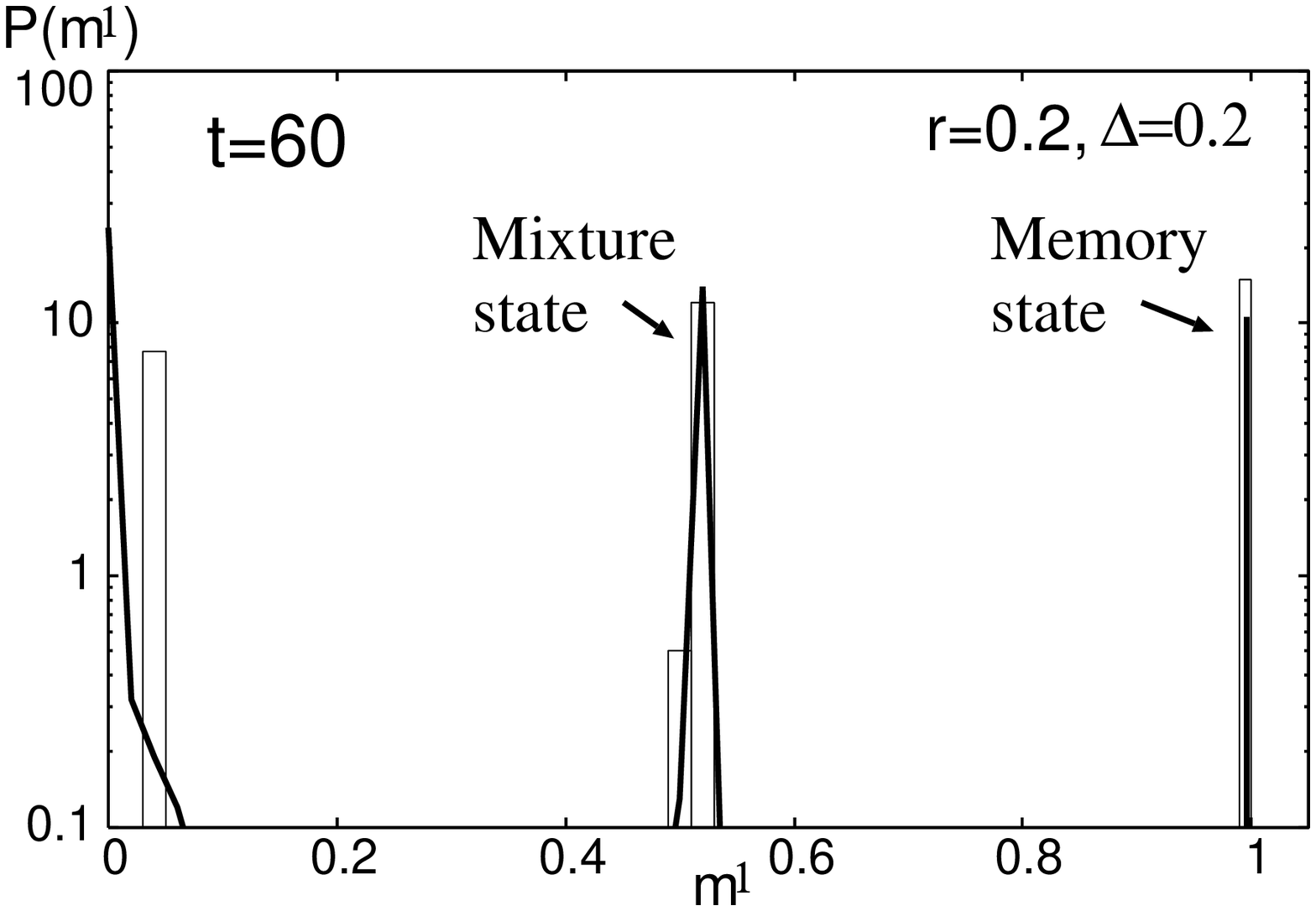}
  \hfill\mbox{}

  \hfill(a)~$t=51$\hfill\hfill(b)~$t=60$\hfill\mbox{}
 \end{center}
 \caption{Distribution of overlaps, $p\left(m^1_t\right)$, at 
 \rev{(a) time $t=51$, and (b) $t=60$ in Fig.~\ref{fig:ovlp_d0.5-0}}.
 \label{fig:hist_d0.5-0}}
\end{figure}

Next, to verify that the state arrives at the mixture state, we consider
a variation of the model that eliminates correlated noise after time
$t=50$. Figure~\ref{fig:ovlp_d0.5-0} shows time evolution of the
overlap, where $\delta=0.5$ for $t\leq50$ and $\delta=0$ for $t>50$. We
find that the state wandering around the mixture state converges on the
mixture state.
Since the overlaps are distributed, we evaluate the marginal PDF of the
overlap $m^1_t$ and compare the theoretical results with those obtained
using computer simulations.  Figure~\ref{fig:hist_d0.5-0} shows the
distribution of the overlap at time $t=51$ and $60$. Abscissas denote
the overlap $m^{1}_t$, and ordinates denote the marginal PDF
$p\left(m^{1}_t\right)$ on a logarithmic scale.  Solid lines denote the
marginal PDF obtained theoretically, and boxes denote histograms obtained
using computer simulations. 
They show positive agreement. We find 
that the states are wandering around either the memory or mixture state.

\section{conclusion}
\label{sec:conclusion}

To analyze the functional roles of correlated noise, we investigated
the associative memory models to which independent and correlated noises
are introduced. In this paper, we analyzed two
types of associative memory models: one with auto- and weak
cross-correlation connections and one with hierarchically correlated
patterns. In the former, we showed that correlated noise can switch
from autoassociative to sequential associative memory. 
Switching memory systems is important for information
processing. The most important point is that correlated noise is not
random, but is added to all neurons mutually. In the latter,
both the memory states and their mixture states are stable. Using only
independent noise, the stochastic transition from the memory to
the mixture state cannot be invoked, but sample dependence does
appear and the stochastic transition can be invoked using 
correlated noise.

In the case where sample dependence appears, the distribution of the
overlaps needed to be analyzed to describe the macroscopic state. We,
therefore, derived the macroscopic dynamic description as a recurrence
relation form of probability density functions. The distributions
obtained theoretically agree with those obtained using computer
simulations.
A transition between attractors cannot be invoked by thermal independent
noise, but can be by synchronous spikes, indicated by Aoyagi and
Aoki\cite{AoyagiAoki2004,AokiAoyagi2006}.  This was demonstrated by our
correlated external input, i.e., correlated noises.  
\rev{In this paper, we examined the cases when $p=3$ for the former, and
$p=1$ and $k=3$ for the latter. When $p\sim{\cal O}(1)$ and $k\sim{\cal
O}(1)$, the behaviors for different values of $p$ and $k$ are
qualitatively same as at present excepting stability of attractors.
}


\section*{Acknowledgment}
This work was partially supported by 
a Grant-in-Aid for Scientific Research on Priority Areas No.~18020007
and No.~18079003, 
a Grant-in-Aid for Scientific Research (C) No.~16500093, 
and a Grant-in-Aid for Young Scientists (B) No.~16700210.

The computer simulation results were obtained using the PC cluster
system at Yamaguchi University.


\end{document}